\documentclass[amsmath,amssymb,prb,reprint,superscriptaddress]{revtex4-1}

\usepackage{graphicx}
\usepackage{epsfig}
\usepackage{rotating}
\usepackage{dcolumn}
\usepackage{bm}
\usepackage{mathrsfs}
\usepackage{leftidx}

\begin{document}


\title{Spin cluster operator theory for the Kagome lattice antiferromagnet}


\author{Kyusung Hwang}
\affiliation{Center for Strongly Correlated Materials Research, Seoul National University, Seoul 151-747, Korea}
\affiliation{School of Physics, Korea Institute for Advanced Study, Seoul 130-722, Korea}

\author{Yong Baek Kim}
\affiliation{School of Physics, Korea Institute for Advanced Study, Seoul 130-722, Korea}
\affiliation{Department of Physics, University of Toronto, Toronto, Ontario M5S 1A7, Canada}

\author{Jaejun Yu}
\affiliation{Center for Strongly Correlated Materials Research, Seoul National University, Seoul 151-747, Korea}

\author{Kwon Park}
\email{kpark@kias.re.kr}
\affiliation{School of Physics, Korea Institute for Advanced Study, Seoul 130-722, Korea}


\date{\today}

\begin{abstract}
The spin-1/2 quantum antiferromagnet on the Kagome lattice provides a quintessential example in the strongly correlated electron physics where both effects of geometric frustration and quantum fluctuation are pushed to their limit.  
Among possible non-magnetic ground states, the valence bond solid (VBS) with a 36-site unit cell is one of the most 
promising candidates.
A natural theoretical framework for the analysis of such VBS order is to  consider quantum states on a bond 
connecting the nearest-neighboring sites as fundamental quantum modes of the system and treat them as effectively independent ``bond particles.''
While correctly describing the VBS order in the ground state,  
this approach, known as the bond operator theory, significantly overestimates the lowest spin excitation energy.
To overcome this problem, we take a next logical step in this paper to improve the bond operator theory
and consider extended spin clusters as fundamental building blocks of the system.
Depending on two possible configurations of the VBS order, various spin clusters are considered: 
(i) in the VBS order with staggered hexagonal resonance, we consider one spin cluster for a David star and two spin clusters with each composed of a perfect hexagon and three attached dimers, and 
(ii) in the VBS order with uniform hexagonal resonance, one spin cluster composed of a David star and three attached dimers. 
It is shown that the majority of low-energy spin excitations are nearly or perfectly flat in energy.
With most of its weight coming from the David star, the lowest spin excitation has a gap much lower than the previous value obtained by the bond operator theory, narrowing the difference against exact diagonalization results.
Despite the better agreement, however, 
the fact that the lowest spin excitation mostly comes from the David star structure is at odds with the bond operator theory as well as previous series expansion studies.
Expecting that the true low-energy spin excitations have contributions from both the David star and its surroundings, we propose to use two super spin clusters in future work treating spin excitations from the both contributions on a more equal footing.   
\end{abstract}


\maketitle

\section{INTRODUCTION\label{sec:INTRODUCTION}}

Quantum antiferromagnet in two spatial dimensions has been a subject of intense research for multiple reasons.
On one hand, the antiferromagnetic order parameter itself does not commute with the antiferromagnetic exchange interaction in the Hamiltonian so that there exist inherent quantum fluctuations in the system.
On the other hand, effects of quantum fluctuations are amplified as the size of the spin moment becomes smaller and the spatial dimension of the system gets lower. 
The antiferromagnetic spin-1/2 chain is the limiting situation where both reach their respective minimum possible values.
Indeed, the antiferromagnetic spin-1/2 chain demonstrates the strongest effect of the quantum fluctuation, which makes the ground state a gapless spin-liquid.\cite{Lieb_Schultz_Mattis,Cloizeaux_Pearson} 
While interesting in many aspects, the one-dimensional quantum antiferromagnet is too much susceptible to quantum fluctuations that it hardly displays any long-range order.  

In this context, the two-dimensional quantum antiferromagnet gives rise to perhaps the most interesting regime where the strength of quantum fluctuation is just so right that there is a delicate balance between order and disorder.
On simple lattices, magnetically ordered states are often found to be energetically favorable.  
For example, the two-dimensional quantum antiferromagnet becomes ordered in the square lattice with staggered spin moments. The ground state in this case is known as the N\'{e}el state.
Realizing quantum disordered states in two spatial dimensions requires a delicate way of tilting the balance between order and disorder.
Geometric frustration is often envisioned as an effective way to do just that.

\begin{figure*}
 \centering
 \includegraphics[width=0.8\linewidth]{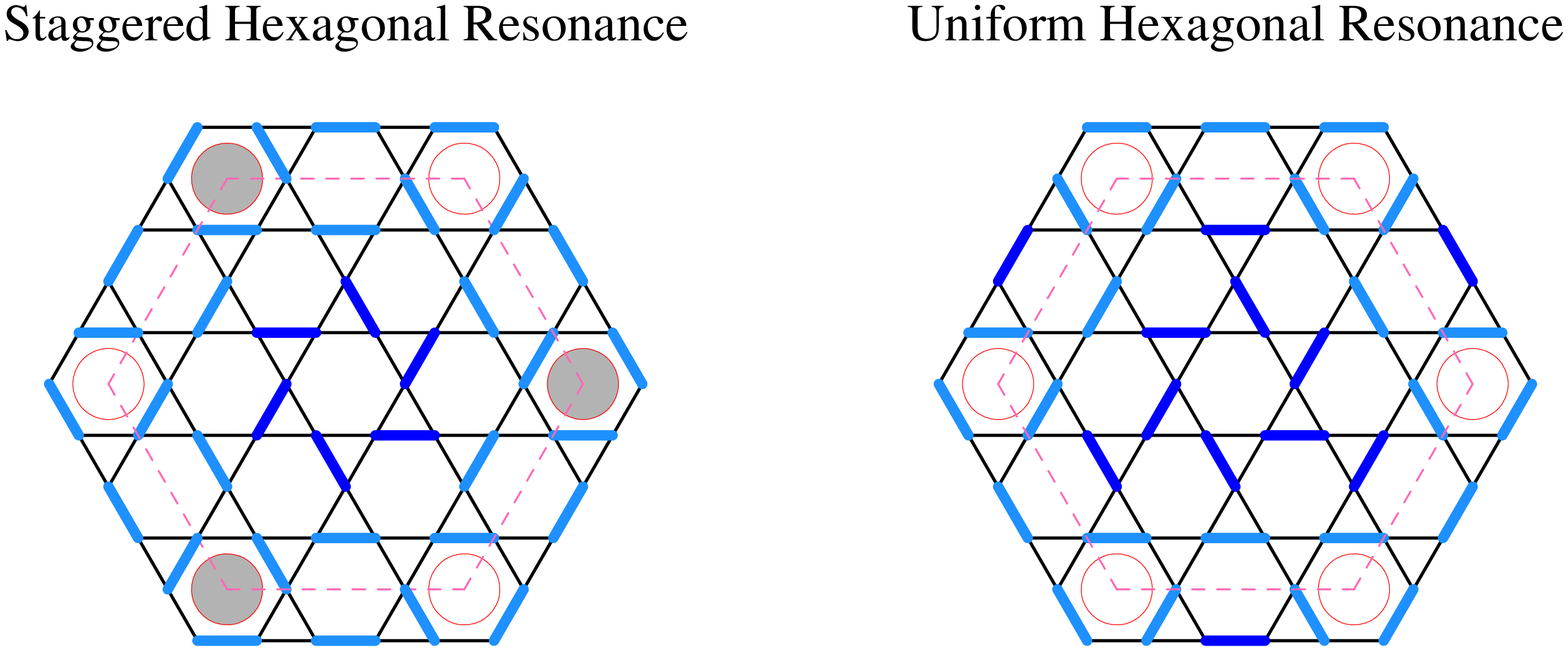}
 \caption{(Color online)
Two spin-singlet covering configurations of the valence bond solid (VBS) ground state with a 36-site unit cell. 
The left figure shows the VBS with staggered hexagonal resonance and the right figure shows the one with uniform hexagonal resonance.
Each dimer, denoted as a thick line in the figure, indicates the position of a spin-singlet bond.
The unit cell with 36 sites (or 18 dimers) is indicated by the enclosing dashed line.
Note that the two VBS order states have different symmetries. 
With respect to the rotation around the center of the pinwheel structure, the VBS order with staggered hexagonal resonance has a sixfold rotation symmetry while that with uniform hexagonal resonance is threefold.
 \label{fig:36_site_unit_cell}}
\end{figure*}

The Kagome lattice is regarded as one of the most geometrically frustrated systems in two dimensions owing to the corner-sharing triangular structure,   
and therefore expected to be the best place to look for the quantum disordered state in spatial dimensions higher than one. 
Indeed, in a recent experimental discovery of an ``ideal" Kagome lattice realized in Herbertsmithites, ZnCu$_{3}$(OH)$_{6}$Cl$_{2}$,~\cite{herbertsmithites_Helton,herbertsmithites_Mendels,herbertsmithites_Ofer} it is shown that the material remains magnetically disordered down to 50 mK while its Curie-Weiss temperature, a measure of the spin exchange energy, is about 300 K.
This experimental finding strongly suggests that the ground state might be spin disordered even at zero temperature, providing a concrete example of fully quantum disordered states in two spatial dimensions.  
Despite this strong suggestion for the existence, however, the precise nature of the spin disordered state is still unknown.

To construct a precise theory for the spin disordered ground state in the Kagome lattice antiferromagnet, it is crucial to use lessons obtained from numerical studies.      
Following are some of the most important lessons from exact diagonalization.~\cite{36-site_ED_Leung_Elser,36_ED_first_excited,ED_Sindzingre} 
(i) The ground state is magnetically disordered.
(ii) The spin-0 excitation seems to have a very small gap or is completely gapless.  
(iii) The spin-1 excitation has a small, but finite gap, which is $0.164 J$  in the $N=36$ system and estimated to be around $0.05-0.1 J$ in the thermodynamic limit. Here, $J$ is the spin exchange energy.

There are two possible candidates for the spin disordered ground state.~\cite{
Zeng_Elser,
Marston_Zeng,
spin_liquid_Sachdev,
36-site_ED_Leung_Elser,
36_ED_first_excited,
Mila,
Sindzingre,
spin_liquid_Hastings,
Nikolic_Senthil,
spin_liquid_Ran,
Singh_Huse_1,
Singh_Huse_2,
Kagome_BOT,
ED_Sindzingre,
Entanglement_Renormalization_Vidal,
quantum_dimer_model_Poilblanc,
JiangWengSheng,
YanHuseWhite,
ED_Sorensen}
One is the spin liquid state where spin moments are disordered without breaking any lattice translation symmetry.~\cite{spin_liquid_Sachdev,spin_liquid_Hastings,spin_liquid_Ran} 
There are two variations within the spin liquid scenario, differing in whether the spin excitation is gapped or gapless.
Considering the results from exact diagonalization and recent density matrix renormalization group (DMRG) studies~\cite{JiangWengSheng,YanHuseWhite}, a gapped version 
of the spin liquid has a potential to be the true ground state. 
The other candidate is the so-called valence bond solid (VBS) order where spins are paired up with one of its nearest neighbors to form a spin singlet, consequently breaking the lattice translation symmetry.~\cite{Marston_Zeng,Nikolic_Senthil,Singh_Huse_1,Singh_Huse_2,Kagome_BOT,quantum_dimer_model_Poilblanc} 
While earlier exact diagonalization as well as series expansion studies favor the VBS order, a recent study using the DMRG method~\cite{YanHuseWhite} seems to indicate that the spin liquid may be closer to the true ground state. 

Resolving the issue as to which state is really the true ground state for the pure Heisenberg model is highly non-trivial and numerically subtle since many competing ground states are very close in energy.  
Setting aside this issue, however, what is important is the fact that, due to the very proximity between various competing ground states, even a small deviation from the pure Heisenberg model Hamiltonian can cause significant changes in the ground state property.
For example, it is shown in the same DMRG study mentioned above~\cite{YanHuseWhite} that a slight reduction in $J$ by less than 2\% in a particular pattern can in fact stabilize the VBS order with a 36-site unit cell rather than the spin liquid.
In addition, there are various important experimental factors that modify the Hamiltonian and thus play crucial roles in determining the true ground state.
Such factors include the Dzyaloshinskii-Moriya interaction, the lattice deformation, the presence of impurity spins, and so on.  
In this context, it is constructive to conduct a comprehensive investigation of all possible spin disordered states and study
their properties.

In this paper, we focus on the VBS state with a 36-site unit cell illustrated in Fig.~\ref{fig:36_site_unit_cell}.  
Spin singlet bonds, denoted as thick lines in the figure, represent the VBS ordering pattern with the 36-site unit cell;
this is believed to be the most favorable configuration by many of the previous studies.~\cite{Marston_Zeng,Nikolic_Senthil,Singh_Huse_1,Kagome_BOT} 
Schematically speaking, the most favorable VBS order is the one with the honeycomb structure of perfect hexagons which in turn surround the pinwheel structure in the David-star shaped region at the core.  
Here the perfect hexagon means that the sides of a hexagon is alternatively covered with three spin-singlet pairs. 
Fig.~\ref{fig:36_site_unit_cell} shows two possible honeycomb arrangements of the perfect hexagons with one having the staggered hexagonal resonance and the other the uniform hexagonal resonance.
In the VBS order with staggered hexagonal resonance, perfect hexagons have an alternating dimer covering pattern (denoted as filled and empty circles inside the hexagon) encircling the pinwheel structure. 
Meanwhile, in the VBS order with uniform hexagonal resonance, all perfect hexagons have the same dimer covering pattern. 
Note that the pinwheel structure itself can have two different orientations, which have exactly the same energy.
The degeneracy between two such states is not lifted even after the pinwheel is coupled with surrounding dimers.   
This is in fact the origin of the gapless spin-0 excitations in the VBS scenario. 
We consider both the VBS order with staggered and uniform hexagonal resonance in this paper since they are equally good candidates for the most stable VBS order.

The VBS state with uniform hexagonal resonance was previously studied by some of the authors within the bond operator theory (BOT).~\cite{Kagome_BOT}
The bond operator theory provides a natural theoretical framework for the analysis of the valence bond solid state by regarding quantum states on a bond connecting the nearest-neighboring sites as fundamental quantum modes of the system.   
At the mean-field level, such quantum modes are treated as effectively independent ``bond particles.''
While correctly producing the VBS order in the ground state, however, the bond operator theory turns out to be quantitatively inaccurate.
In particular, the spin-1 excitation gap is predicted to be much larger than that obtained from exact diagonalization of the 36-site cluster while the ground state energy is not too bad considering the level of approximation. 

In addition, a peculiar property of the bond operator theory is that the lowest spin-1 excitations entirely originate from the surroundings [light blue (gray) in the right figure of Fig.~\ref{fig:36_site_unit_cell}] that encircle the core pinwheel structure. 
This is a consequence of the bond operator theory that core dimers [dark blue (gray) in the right figure of Fig.~\ref{fig:36_site_unit_cell}] are completely decoupled from the surrounding ones.
While this conclusion may be fully consistent with series expansion results,
the large discrepancy in the lowest spin-excitation gap energy leaves the accuracy of the bond operator analysis in question. 
Moreover, the bond operator theory predicts that the spin-1 excitations from the pinwheel structure are all degenerate with the energy gap being exactly $1 J$. This is clearly not correct.

While the problem certainly arises from the coarse nature of the mean-field approximation, 
there is no systematic way to improve the mean-field theory, especially for the ``hard-core'' constraint imposing that no two bond particles can occupy the same dimer site. 
In this paper, 
instead of looking for a perturbative way to include higher-order diagrams,  
we set out to improve the bond operator theory by extending the fundamental building block of the system from a bond to a more extended cluster of spins.  

\begin{figure}[b]
 \centering
 \includegraphics[width=0.8\linewidth]{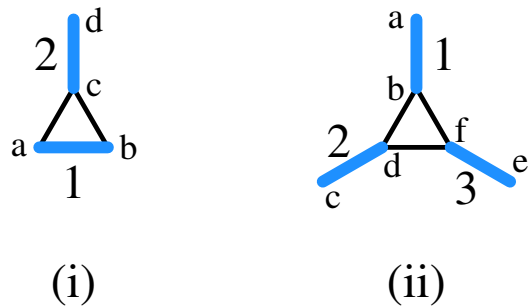}
 \caption{(Color online)
Two different structures of the inter-dimer interaction:
(i) topologically orthogonal and (ii) empty triangle structure.
Alphabetical and numerical indices denote the individual spin and the dimer, respectively. 
 \label{fig:inter_dimer_int}}
\end{figure}

To determine which spin clusters are most appropriate, 
it is worthwhile to investigate the nature of the inter-dimer interaction carefully.  
To this end, it is important to note that all inter-dimer interactions can be classified into two basic units according to their geometric structures:
(i) topologically orthogonal and (ii) empty triangle structure. 
See Fig. \ref{fig:inter_dimer_int} for illustration.
Extended spin clusters can be constructed by the combination of these two basic units.
Next, by using exact diagonalization, we investigate how each basic structure affects the low-energy spectrum of the extended spin clusters. 
As a result, it is shown that the empty triangle structure mainly contributes to the lowering of the ground state energy while the topologically orthogonal structure stabilizes the ground state itself by reducing quantum fluctuations about the VBS order.

We first test the bond operator theory as to how well it captures the effects of the topologically orthogonal as well as the empty triangle structure.
The test shows that the bond operator mean-field theory is quite accurate for capturing the effects of the empty triangle structure while it is not so in the case of the topologically orthogonal structure.
Based on this result, we conclude that the most appropriate spin cluster is the one containing the maximal number of the topologically orthogonal structures allowed by the VBS order. An important assumption here is that, similar to the bond operator mean-field theory, the effects of all remaining empty triangle structures are well captured by the mean-field theory formulated in terms of the quantum modes formed in the extended spin clusters.
For convenience, let us call this approach the spin cluster operator theory (SCOT).

\begin{figure*}
  \includegraphics[width=1.\textwidth]{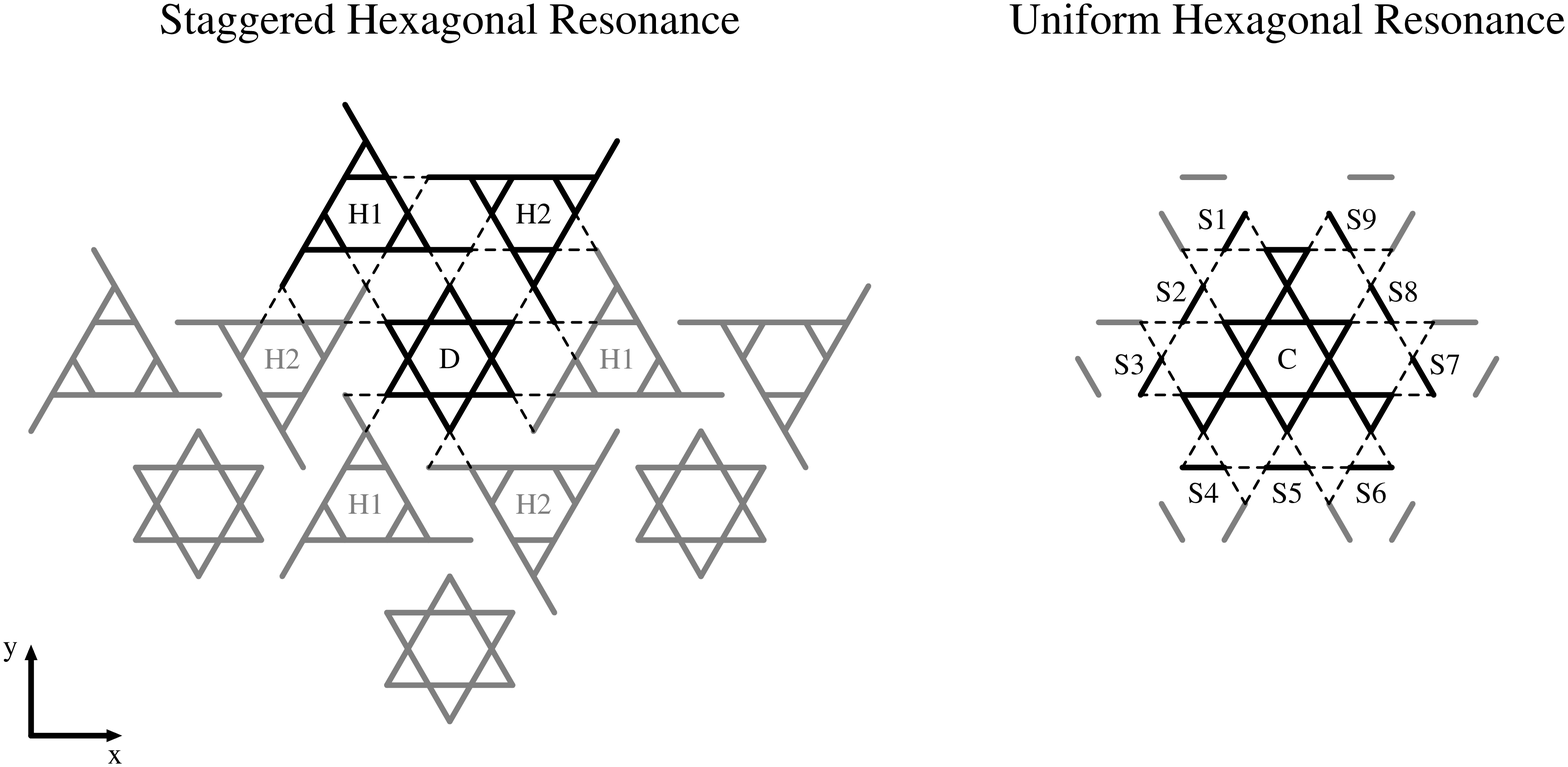}
  \caption{
  Spin cluster covering for the VBS order with staggered hexagonal resonance (left figure) and that with uniform hexagonal resonance (right figure).  
  The unit cell of the VBS order with staggered hexagonal resonance consists of three spin clusters that include a David star, D, and two extended hexagons, H1 and H2.
  Similarly, the unit cell of the VBS order with uniform hexagonal resonance is composed of a core cluster, C, and nine surrounding dimers, S1 - S9. 
  \label{fig:Cluster covering}}
\end{figure*}

Figure~\ref{fig:Cluster covering} shows spin cluster coverings for each type of the VBS order depicted in Fig.~\ref{fig:36_site_unit_cell}.
As one can see,
the VBS order with staggered hexagonal resonance is covered with one spin cluster for a David-star structure (denoted as D) and two spin clusters with each composed of a perfect hexagon and three attached dimers (denoted as H1 and H2).
Meanwhile, the VBS order with uniform hexagonal resonance is covered with one spin cluster composed of a David-star and three attached dimers (denoted as C) and nine dimers for spin-singlets (denoted as S1, $\cdots$, S9). 
Here spin clusters are chosen according to the following three conditions:
(i) the VBS order formed inside each individual spin cluster should be consistent with that of the whole lattice, 
(ii) the shape of spin clusters should preserve the symmetry of the VBS order, and
(iii) spin clusters should include as many inter-dimer interactions with the topologically orthogonal structure as possible within themselves. 
There is one additional practical constraint in the choice of the above spin clusters, which comes from the technical problem that the size of the Hilbert space grows exponentially as the number of spins increases. 
For this reason, in this study, we choose an appropriate subset of the unit cell as a spin cluster. 
For future work, we propose to investigate the ``super cluster'' that contains all topologically orthogonal structures within the whole unit cell.
Details of our proposal are discussed later in this paper.

The ground state energy as well as the spin-1 excitation energy spectrum
can be computed via mean-field analysis of the spin cluster operator theory. 
For the VBS order with staggered hexagonal resonance, the ground state energy is given by $-0.428 J $ per spin which is not too bad in comparison with the 36-site exact diagonalization result, $-0.438 J$, considering the coarse nature of the mean-field theory.
The spin-1 excitation spectrum is found to be almost flat with very narrow band widths in the low energy sector.
The lowest spin-1 excitation occurs at the K point in the first Brillouin zone with the energy gap being $0.323 J$.
Compared to the mean-field result of the bond operator theory for the same VBS order, $0.833 J$,
the above spin-1 excitation energy gap is in much better agreement with exact diagonalization.
For the VBS order with uniform hexagonal resonance, the core cluster, denoted as C in Fig.~\ref{fig:Cluster covering}, is completely decoupled from the surrounding dimers, resulting in exactly the same ground state energy as that of the bond operator theory, $-0.422 J$ per site.
Spin-1 excitations consist of two entirely separate groups with one coming from the core cluster and the other from the surrounding dimers. 
The lowest spin-1 excitation occurs from the core cluster with its energy gap being $0.243 J$.
This energy gap is also much reduced from the BOT result for the same VBS order, $0.776 J$.

Despite these improvements, however, the spin-1 excitation energy gap is still somewhat higher than the exact diagonalization result.  
Also, an important feature of the low energy excitations is that they predominantly originate from the triplet states within the pinwheel structure in Fig.~\ref{fig:Cluster covering}. 
This fact is at odds with the bond operator theory as well as previous series expansion studies.
In combination with the reduced, but still larger size of the spin gap, this suggests that the effects of the topologically orthogonal inter-dimer interaction structures are not fully incorporated.  
Expecting that the true low-energy spin excitations have contributions from both the pinwheel and its surroundings, we propose to use super spin clusters in future work, which treat spin excitations from the both contributions on a more equal footing.
The proposed super clusters are chosen so that they include all topologically orthogonal inter-dimer structures within the 36-site unit cell. 
Details of the proposal are discussed in the Discussion section.

The rest of the paper is organized as follows. 
In Sec.~\ref{sec:MOTIVATION}, we study
physical effects of the inter-dimer interaction in two basic units with different geometric structures: (i) topologically orthogonal and (ii) empty triangle structures. 
The bond operator theory is tested on such basic unit structures in order to clarify how well it captures the effects of each inter-dimer interaction.
The spin cluster operator theory is proposed as a solution to the problems emerging in the bond operator theory.
In Sec.~\ref{sec:SCOT_I} and \ref{sec:SCOT_II}, the formalism for the spin cluster operator theory is constructed for both the VBS order with staggered and uniform hexagonal resonance. 
Results of the spin cluster operator theory are found in Sec.~\ref{sec:RESULTS_I} and \ref{sec:RESULTS_II}. 
In Sec.~\ref{sec:DISCUSSION}, we propose two super clusters of spins; the analysis of which can further improve the accuracy of the spin cluster operator theory. 
Finally, we conclude in Sec.~\ref{sec:SUMMARY}.
\\
\\

\section{MOTIVATION\label{sec:MOTIVATION}}

\subsection{Topologically orthogonal versus empty triangle inter-dimer interaction
\label{subsec:topologically_orthogonal_vs_empty_triangle}}

As mentioned in the Introduction, in this paper, the valence bond solid state with a 36-site unit cell is regarded as a candidate ground state for the Kagome-lattice antiferromagnetic Heisenberg model (KLAHM):
\begin{eqnarray}
H= J \sum_{\left<i,j\right>} {\bf S}_i \cdot {\bf S}_j \;,
 \label{eq:Hamiltonian}
\end{eqnarray}
where $\left< i,j \right>$ denotes that $i$ and $j$ are the nearest neighbors.
Here, the spin exchange energy, $J$, is positive since we are interested in  the antiferromagnetic model. 
The VBS order with a 36-site unit cell has a dimer covering pattern represented by a honeycomb array of perfect hexagons with a pinwheel structure at the center of each cell.   
This particular pattern of the VBS order is generally believed to be the most favorable among various possibilities 
while the underlying physical reason why this is so varies from theory to theory. 

With the VBS order composed of the two structures, 
(i) a honeycomb array of perfect hexagons and (ii) a pinwheel structure at the center of each honeycomb cell, 
different viewpoints occur regarding which structure plays a more important role for the stability of the VBS order.  
By using a duality mapping to the fully frustrated Ising model,
Nikolic and Senthil~\cite{Nikolic_Senthil} advocated a viewpoint that the most preferred dimer covering is the one simultaneously maximizing the number of perfect hexagons and pinwheels (or the 12-bond star-shaped flippable loops as called in the paper).
Another viewpoint is provided by Singh and Huse~\cite{Singh_Huse_1} who, by using series expansion, argue that the ground state energy is minimized when every three empty triangles are bound into perfect hexagons. 
Such formed perfect hexagons are furthermore arranged in such a way that empty triangles between the nearest perfect hexagons share as many dimer bonds as possible. 
The resulting dimer covering is exactly the same honeycomb array of perfect hexagons as before.
In this picture, the pinwheel structure at the center of the unit cell is just a simple byproduct of the overall covering pattern.
Some of the authors in this paper proposed a yet different viewpoint that the number of topologically orthogonal inter-dimer structures is maximized for the stability of the dimer covering pattern.~\cite{Kagome_BOT}
Both the honeycomb array of perfect hexagons and the pinwheel structure at the center of the unit cell are the direct consequence of such maximization.
In this paper, we take this approach.

To reveal why this approach is physically appealing, let us investigate how the two different types of the basic inter-dimer interaction unit, (i) topologically orthogonal and (ii) empty triangle structure, affect the ground state as well as the low-energy excitations.  
We first show exact diagonalization results for each basic inter-dimer interaction unit in Table~\ref{tab:inter_dimer_int_ED}.    
Then, the results for further complicated spin clusters extending the two basic units are presented in Table~\ref{tab:cluster_series_ED} and \ref{tab:nonempty_empty_combination}.

\begin{table}
 \centering
  \includegraphics[width=1.0\linewidth]{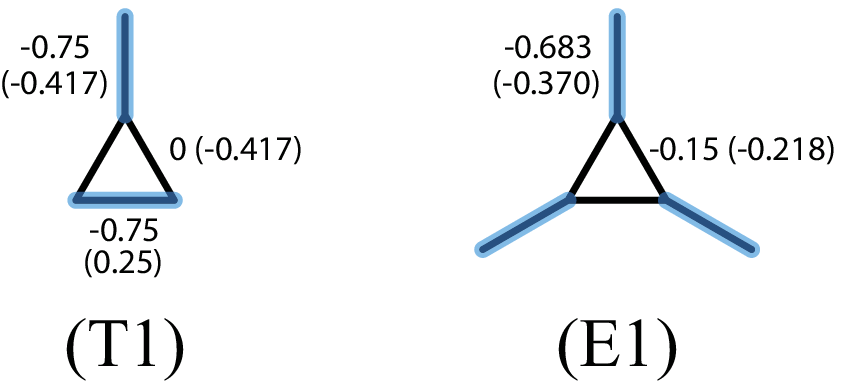}
  \begin{ruledtabular}
  \begin{tabular}{cccc}
  & 
  $\epsilon_{gr}/J$ 
  & $\Delta_{S=1}/J$
  & $N_{1J}$
  \\
  \hline
  (T1) &-0.375 &0.5 &7
  \\
  \hline
  (E1) &-0.417 &0.736 &7
  \\
  \end{tabular}
  \end{ruledtabular}
  \caption{(Color online) Two basic units for the inter-dimer interaction: (T1) topologically orthogonal and (E1) empty triangle structure. 
  Thick (blue) line segments denote where spin singlet dimers are located.
  The table shows exact diagonalizaion results for the ground state energy per spin, $\epsilon_{gr}/J$, the lowest spin-1 excitation energy gap, $\Delta_{S=1}/J$, and the number of states within $1J$ from the ground state in energy, $N_{1J}$.  
  Numbers near thick and thin lines in the figure denote the spin-spin correlation value, $\left< {\bf S}_i \cdot {\bf S}_j \right>$, for the ground state. Corresponding numbers for the lowest spin-1 excitation are given in parenthesis.
  \label{tab:inter_dimer_int_ED}}
\end{table}

The most important feature of the spin-singlet ground state in the topologically orthogonal unit structure, denoted as T1, is that the exact ground state is a simple direct product of the two spin-singlet states formed in each dimer bond depicted as thick (blue) lines in Table~\ref{tab:inter_dimer_int_ED}.  
Considering that such direct product state is simply the ground state with the inter-dimer interaction completely turned off,
it means that there is absolutely no quantum fluctuation about the valence bond solid order. 
In fact, no matter how large the spin cluster may become, the direct product state of spin singlets can be shown to be an exact energy eigenstates of the Heisenberg-model Hamiltonian (while not necessarily the ground state) so long as all inter-dimer connections are topologically orthogonal.
The Shastry-Sutherland lattice provides a realistic example where actually all inter-dimer connections are topologically orthogonal in the thermodynamic limit.\cite{ShastrySutherland}
In this situation, it is known that the VBS state remains as the exact ground state so long as the inter-dimer spin exchange energy is less than a certain critical percentage (estimated to be around 0.7) of the intra-dimer counterpart.\cite{MiyaharaUeda}
It is worthwhile to mention that an actual material, SrCu$_2$(BO$_3$)$_2$, has been discovered to be a faithful realization of the Shastry-Sutherland lattice with the VBS state being the ground state.\cite{Kageyama} 
Coming back to the Kagome lattice problem, what is important here is that the VBS order can be stable in the topologically orthogonal inter-dimer interaction structures owing to the absence of the quantum fluctuation.

In the case of the empty triangle unit structure, denoted as E1 in Table~\ref{tab:inter_dimer_int_ED}, the ground state wave function is no longer a direct product of spin singlet wave functions. 
The overall characteristics of the ground state is, however, still quite well captured by the VBS order with three spin singlets formed in each dangling bond attached to the empty triangle. 
Here, we use a simple criterion that the closer the spin-spin correlation value, $\left< {\bf S}_i \cdot {\bf S}_j \right>$, is to -0.75 (0.25), the better the state is described by the spin singlet (triplet) wave function. 
Meanwhile, the ground state energy per spin in the empty triangle unit structure, $-0.417 J$, is much lower than that of the topologically orthogonal counterpart, $-0.375 J$.
In combination with the loss of the exact VBS order, 
this fact can be interpreted as if the empty triangle structure allows more quantum fluctuations in exchange for a lowered ground state energy.
So far, the conclusion is that the topologically orthogonal structure stabilizes the VBS order by removing quantum fluctuations while the empty triangle structure tries to lower the ground energy by allowing some. 
It is shown below that this conclusion remains valid for spin clusters larger than the two basic structures studied here.

\begin{table*}
 \centering
 \includegraphics[width=0.9\linewidth]{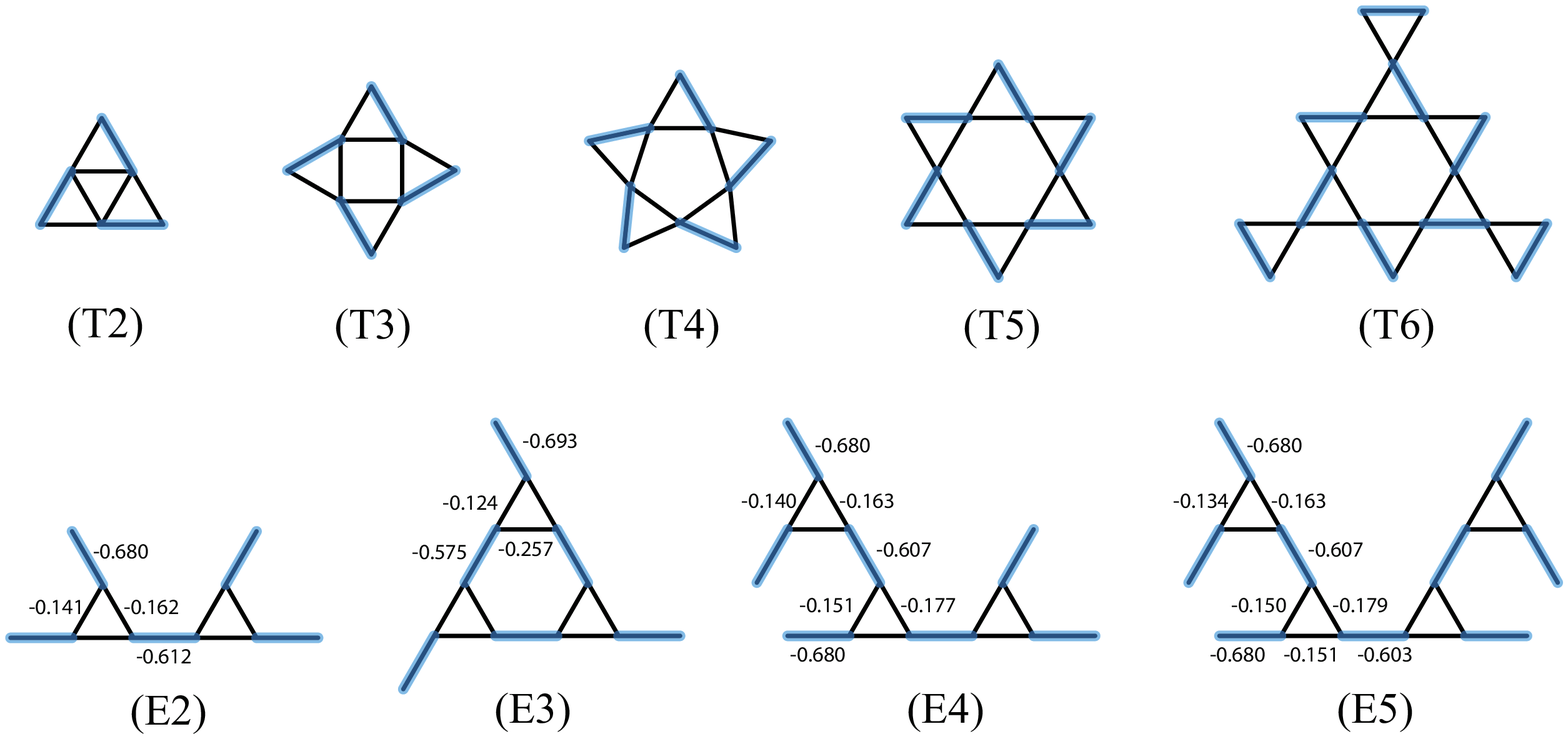}
 \begin{minipage}{0.5\linewidth}
  \centering%
  \includegraphics[angle=270,width=1.0\linewidth]{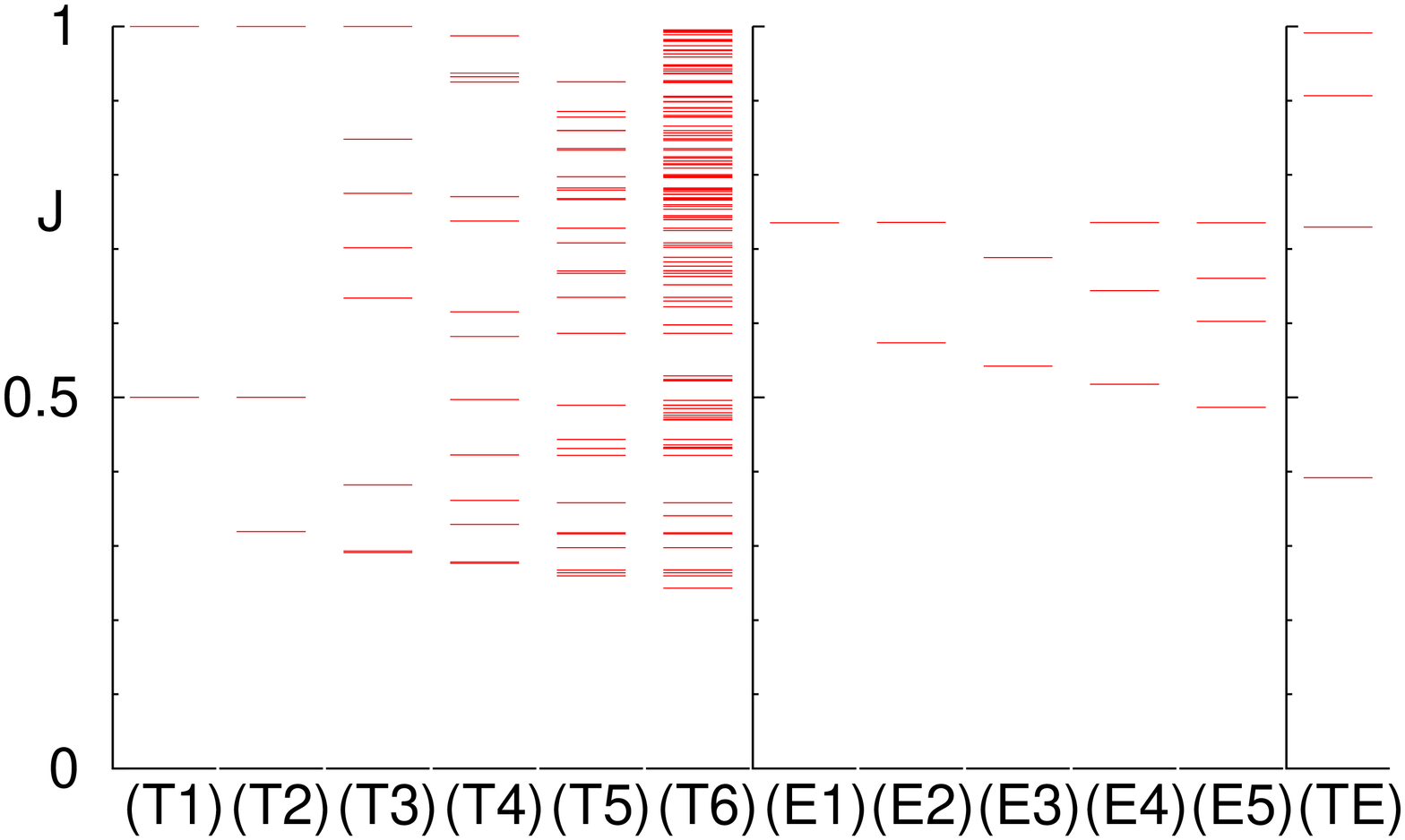}
 \end{minipage}
 \begin{minipage}{0.45\linewidth}
 \begin{ruledtabular}
 \begin{tabular}{cccc}
  &
  $\epsilon_{gr}/J$
  &
  $\Delta_{S=1}/J $\footnote{In spin clusters studied, the lowest spin excitations are all spin triplets.} 
  &
  $\textup{N}_{1J}$
  \\
  \hline
  (T2) &-0.375 &0.319 &11
  \\
  \hline
  (T3) &-0.375 &0.291 &31
  \\
  \hline
  (T4) &-0.375 &0.277 &70
  \\
  \hline
  (T5) &-0.375 &0.260 &122
  \\
  \hline
  (T6) &-0.375 &0.243 &844
  \\
  \hline
  \hline
  (E2) &-0.426 &0.574 &10
  \\
  \hline
  (E3) &-0.443 &0.542 &10
  \\
  \hline
  (E4) &-0.430 &0.518 &13
  \\
  \hline
  (E5) &-0.432 &0.487 &16
  \\
  \hline
  \hline
  (TE)  &-0.408 &0.392 &11
 \end{tabular}
 \end{ruledtabular}
 \end{minipage}
 \begin{minipage}{1.0\linewidth}
  \caption{(Color online)
  Schematic diagram for the spin clusters containing only the topologically orthogonal inter-dimer interactions, T2-T6, and the empty triangle interactions, E2-E5. 
  Dimers (light blue) indicate the valence bonds formed in the ground state.
  In the spin clusters T2-T6 where entire dimer connections are topologically orthogonal, the ground state wave function is precisely given by a direct product of the spin-singlet wave functions formed in each dimer. 
  In this situation, the spin-spin correlation value, $\left< {\bf S}_i \cdot {\bf S}_j \right>$, becomes exactly -0.75 within dimers and 0 between dimers.  
  It is important to note that each dimer configuration in T2-T6 is one of the twofold degenerate ground states.
  The other state not shown in the figure is obtained by reversing the orientation of the dimer covering pattern in the pinwheel structure.
  In the spin clusters E2-E5 where all inter-dimer interactions are through empty triangle structures, the spin-spin correlation values (shown near each corresponding bond in the figure) of the ground state are overall similar to those of the simple spin-singlet product state while the wave function is not precisely in the product form.   
  {\bf Lower left figure:} energy spectrum of each spin cluster up to $1J$ of the excitation energy.
  In this figure, TE is a new spin cluster containing the both topologically orthogonal and empty triangle structure. 
  See Table~\ref{tab:nonempty_empty_combination} for detail.
  {\bf Lower right table:} the ground state energy per spin, $\epsilon_{gr}/J$, and the spin-1 excitation energy gap, $\Delta_{S=1}/J$. 
  N$_{1J}$ denotes the number of states within $1J$ of the excitation energy from the ground state.
  \label{tab:cluster_series_ED}}
 \end{minipage}
 \begin{minipage}{1.0\linewidth}
 \includegraphics[width=0.15\linewidth]{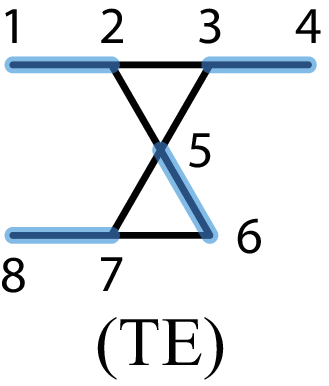}
 \begin{ruledtabular}
 \begin{tabular}{cccccccccccc}
  & $E/J$ & $S$ & $\left< {\bf S}_1 \cdot {\bf S}_2 \right>$ & $\left< {\bf S}_3 \cdot {\bf S}_4 \right>$ & $\left< {\bf S}_5 \cdot {\bf S}_6 \right>$ & $\left< {\bf S}_7 \cdot {\bf S}_8 \right>$ & $\left< {\bf S}_2 \cdot {\bf S}_3 \right>$ & $\left< {\bf S}_3 \cdot {\bf S}_5 \right>$ & $\left< {\bf S}_5 \cdot {\bf S}_2 \right>$ & $\left< {\bf S}_5 \cdot {\bf S}_7 \right>$ & $\left< {\bf S}_6 \cdot {\bf S}_7 \right>$
  \\
  \hline
  Ground state &-3.265 &0 &-0.676 &-0.676 &-0.662 &-0.742 &-0.134 &-0.169 &-0.169 &0.013 &-0.050
  \\
  \hline
  $1^\textrm{st}$ excited state &-2.873 &1 &-0.623 &-0.623 &0.139 &-0.470 &-0.037 &-0.280 &-0.280 &-0.254 &-0.446
  \\
  \hline
  $2^\textrm{nd}$ excited state &-2.535 &1 &-0.216 &-0.216 &-0.633 &-0.729 &-0.448 &-0.111 &-0.111 &-0.007 &-0.064
  \\
  \end{tabular}
  \end{ruledtabular}
  \caption{(Color online)
  Energy, $E/J$, Spin, $S$, and the spin-spin correlation value, $\left< {\bf S}_i \cdot {\bf S}_j \right>$, for the ground, the first-excited and the second-excited state in a spin cluster, TE, containing the both topologically orthogonal and empty triangle structure. 
 \label{tab:nonempty_empty_combination}}
 \end{minipage}
\end{table*}

Table~\ref{tab:cluster_series_ED} shows a variety of spin clusters composed of the topologically orthogonal, T2-T6, and the empty triangle, E2-E5, inter-dimer interaction structures. 
In spin clusters T2-T6 where entire dimer connections are topologically orthogonal, the ground state wave function is exactly given by the direct product of spin-singlet wave functions formed in each dimer.
Similar to the topologically orthogonal unit structure, T1, there is absolutely no quantum fluctuation in the ground state for the VBS order so long as all inter-dimer connections are topologically orthogonal. 
This plays a very important role for the stability of the VBS order in the Kagome-lattice antiferromagnet. 
We believe that, given the huge degeneracy in the spin configuration of the (semi-classical) ground state in the KLAHM, the ground state tries to lift as much degeneracy as possible by minimizing the quantum fluctuation.
One way to achieve this goal is to break the lattice translational symmetry with a 36-site unit cell where the number of topologically orthogonal inter-dimer structures is maximized. 
In view of an effective statistical model obtained via quantum-to-classical mapping, we envision that the effective temperature is less than the critical value so that the system tries to minimize quantum fluctuations.
This viewpoint is similar to that of Nikolic and Senthil.~\cite{Nikolic_Senthil}

Despite the formation of a pure VBS order, the quantum fluctuation is not actually completely gone due to a two-fold degeneracy in the pinwheel structure. 
Note that all topologically orthogonal spin clusters, (T2)-(T6), have a pinwheel-shaped dimer covering structure in the middle.
Each VBS ground state shown in the Table~\ref{tab:cluster_series_ED}
has a degenerate energy partner that can be obtained by reversing the orientation of the dimer covering pattern in the pinwheel structure.
This fact is actually very important in view of the numerical result that the spin-0 excitation is known to be either completely gapless or gapped with a very small gap.  
Reversing the dimer covering orientation in the pinwheel structure changes neither the spin value nor the energy, which, therefore, becomes the most natural candidate for the spin-0 excitation mode.

Now, let us switch gears to spin excitations.
As one can see from the lower left figure in Table~\ref{tab:cluster_series_ED}, the spin-1 excitation gap decreases as the number of topologically orthogonal inter-dimer interactions increases from T1 to T6.   
At the same time, more and more spin-1 excitations are available within $1J$ in energy from the ground state, leading to a dense low-energy spectrum. 
It is shown later that it is this decrease (as well as the 
dense low-energy spectrum) that the bond operator theory misses, eventually leading to a gross overestimation of the spin gap.

In spin clusters E2-E5, all inter-dimer interactions are through the empty triangle.
Similar to E1, the ground state wave function is not given by a simple direct product of spin-singlet wave functions, but the overall spin-spin correlation is still well represented by the VBS order.
To see this, look at the spin-spin correlation values given near each corresponding bond in Table~\ref{tab:cluster_series_ED}.
Along the thick lines representing the valence bonds, the spin-spin correlation values are close to $-0.75 J$ while their magnitude is much reduced along the bonds in the empty triangle structure.
The ground state energy of empty triangle spin clusters is much lower than that of the topologically orthogonal counterparts, $-0.375 J$.
In fact, the lowest energy configuration is obtained in E3 where three empty triangles form a closed circuit resulting in the perfect hexagon.  
As for the excitation spectrum, low-energy spin excitations are quite sparse compared to those of the topologically orthogonal spin clusters. 
This result confirms again that the main role of the empty triangle structure is to lower the ground state energy.

All spin clusters considered so far are entirely composed of the inter-dimer interaction structures in a single type.
In Table~\ref{tab:nonempty_empty_combination}, we investigate what happens when both types coexist.  
The simplest inter-dimer structure containing both structures, TE, is composed of a single topologically orthogonal structure, T1, and a single empty triangle structure, E1, as depicted in Table~\ref{tab:nonempty_empty_combination}.

One of the most important features of TE is that the ground state overall maintains the VBS order of each individual inter-dimer structure. 
That is to say, being the highest in magnitude, the spin-spin correlation value is quite close to -0.75 along (1,2), (3,4), (5,6), and (7,8) link. 
Note that $(i, j)$ stands for the link connecting spin $i$ and $j$.
Meanwhile, the corresponding spin-spin correlation value is very small for the (5,7) and (6,7) link in a close agreement with a single topologically orthogonal unit structure, T1.
Also, along the (2,3), (3,5), and (2,5) link, the spin-spin correlation value is roughly the same as that of a single empty triangle unit structure, E1, $-0.15$. 
Another important feature of TE is that its topologically orthogonal part, T1, has very little contributions to the lowering of the ground state energy.
With the ground state energy per spin being $-0.408 J$, the contribution of T1 alone, $J \left< {\bf S}_5 \cdot {\bf S}_6 + {\bf S}_7 \cdot {\bf S}_ 8+ {\bf S}_5 \cdot {\bf S}_7 + {\bf S}_6 \cdot {\bf S}_7 \right>$, is only $-0.360 J$ per spin.
It is due to the empty triangle structure that the ground state has such a low energy mentioned in the above.

As for the excitations, it is found that the first excited state of TE has  strong characteristics of the lowest-energy excitation in T1 while the second excited state is rather close to that of E1.
To be specific, the first excited state of TE is essentially obtained by breaking the spin singlet bond formed in the (5,6) link into a spin triplet, which can be seen via the change in the spin-spin correlation value from $-0.662$ to $0.139$. 
In a close analogy with the lowest-energy excitation in T1, this spin-singlet breaking in turn spreads the spin-singlet characteristics along the (7,8) link to nearby links of (5,7) and (6,7).  
Similarly, the second excited state of TE resembles the lowest-energy excitation of E1. 
Note that this is consistent with the fact that the empty triangle structure has a higher spin excitation energy gap than the topologically orthogonal counterpart.

To summarize, three lessens are obtained from exact diagonalization of various small spin clusters: 
(i) there is absolutely no quantum fluctuation about the VBS order in the spin clusters entirely composed of the topologically orthogonal structures,
(ii) the empty triangle structure plays an important role for the lowering of the ground state energy, and
(iii) the VBS order is quite robust in the ground state even when two different types of the inter-dimer structures coexist. 
This observation provides a motivation for the VBS order in the Kagome lattice antiferromagnet.
In our picture, the strong geometric frustration in the Kagome lattice prefers the ground state minimizing the quantum fluctuation.
The VBS order achieves this goal by maximizing the number of topologically orthogonal inter-dimer structures with a 36-site unit cell within which quantum fluctuations are completely suppressed. 
This suppression, however, cannot occur in the whole lattice (unless the lattice is the Shastry-Sutherland lattice).
Across the boundary of the unit cell, inter-dimer interactions are in fact solely mediated by the empty triangle structure.
In this situation, the 36-site unit cell should be arranged in such a way that the ground state energy is minimized. 
As mentioned previously, the lowest energy configuration is obtained when three empty triangles are bound into a closed packet, leading to the perfect hexagonal structure.    
Two such final VBS order patterns are considered in this paper.

In Sec.~\ref{sec:SCOT_I} and \ref{sec:SCOT_II}, we present a theoretical framework providing a natural platform for describing the VBS order ground state as well as its excitations. 
Before we do that, it is instructive to test how well the bond operator theory
works for the two inter-dimer interaction unit structures, T1 and E1.
This test sets a stage for the spin cluster operator theory where energy eigenstates emerging from extended spin clusters, rather than those emerging from a single bond, are treated as independent quasi-particles of the system.

\subsection{Testing the bond operator theory\label{subsec:testing_BOT}}

We begin by writing the Heisenberg-model Hamiltonian for a single topologically orthogonal unit structure, T1, shown in Fig.~\ref{fig:inter_dimer_int}:
\begin{equation}
 H_{\textrm{T1},J}
 = J \left( {\bf S}_a \cdot {\bf S}_b + {\bf S}_c \cdot {\bf S}_d \right) 
 + \lambda J \left( {\bf S}_a + {\bf S}_b \right) \cdot {\bf S}_c,
 \label{eq:H_top}
\end{equation}
where $\lambda$ is introduced for computational convenience, but is set to be unity at the end of computation.
In the bond operator theory,~\cite{SachdevBhatt} the Hamiltonian is rewritten in terms of a new class of operators for collective quasi-particles emerging from a bond.

Specifically, the bond operators are defined as follows:
\begin{subequations}
\label{eq:BOT_basis}
\begin{eqnarray}
 && \left| s \right> = s^{\dagger} \left| 0 \right> = \frac{1}{\sqrt{2}} \left( \left| \uparrow \downarrow \right> - \left| \downarrow \uparrow \right> \right) \;,
 \\
 && \left| t_x \right> = t_x^{\dagger} \left| 0 \right> = -\frac{1}{\sqrt{2}} \left( \left| \uparrow \uparrow \right> - \left| \downarrow \downarrow \right> \right) \;,
 \\
 && \left| t_y \right> = t_y^{\dagger} \left| 0 \right> = \frac{i}{\sqrt{2}} \left( \left| \uparrow \uparrow \right> + \left| \downarrow \downarrow \right> \right) \;,
 \\
 && \left| t_z \right> = t_z^{\dagger} \left| 0 \right> = \frac{1}{\sqrt{2}} \left( \left| \uparrow \downarrow \right> + \left| \downarrow \uparrow \right> \right) \;,
\end{eqnarray}
\end{subequations}
where $s^{\dagger}$ and $t_{\alpha}^{\dagger}~(\alpha=x,y,z)$ are the respective operators for creating a spin singlet and triplet state.
While either boson or fermion statistics can be assigned to bond particles as long as the Lie algebra for the physical spin is correctly reproduced,
it is convenient to treat both $s$ and $t_{\alpha}$ as boson operators. 

The spin operators for the left and right site, ${\bf S}_L$ and ${\bf S}_R$, are represented in the bond operator formalism as follows: 
\begin{subequations}
\label{bond_op}
\begin{eqnarray}
 &&S_{L\alpha} = \frac{1}{2} \left( s^{\dagger}t_{\alpha} + t_{\alpha}^{\dagger}s - i\epsilon_{\alpha\beta\gamma}t_{\beta}^{\dagger}t_{\gamma} \right) \;,
 \\
 &&S_{R\alpha} = \frac{1}{2} \left( - s^{\dagger}t_{\alpha} - t_{\alpha}^{\dagger}s - i\epsilon_{\alpha\beta\gamma}t_{\beta}^{\dagger}t_{\gamma} \right) \;,
\end{eqnarray}
\end{subequations}
where $\hbar=1$, $\alpha,\beta,\gamma \in \{x,y,z\}$, and the Einstein summation convention is used. 
While correct, the above representation alone is not yet sufficient to define the physical spin operator.  
What is necessary is the hard-core constraint:
\begin{eqnarray}
 s^{\dagger}s + t_{\alpha}^{\dagger}t_{\alpha} = 1 \;.
 \label{eq:hardcore}
\end{eqnarray}
In other words, the spin operator representation in the above can generate the correct Lie algebra only in the Hilbert space restricted by the hard-core constraint.

The main reason why the bond operator theory is useful is that it provides a convenient framework to describe the VBS order.
The VBS order is represented in the bond operator theory through the condensation of $s$ bosons.
Technically speaking, this means that both the annihilation and creation operator, $s$ and $s^\dagger$, are replaced by a simple number, $\bar{s}$, which denotes the condensation density of spin-singlet pairs.

With everything put together, the final Hamiltonian in the bond operator representation is written as follows:
\begin{equation}
 H_\textrm{T1} = H_{\textrm{T1},J}(\bar{s},t_\alpha) - \sum_{i=1}^{2} \mu_i \left( {\bar s}_{i}^2 + t_{i\alpha}^{\dagger}t_{i\alpha} - 1 \right),
\end{equation}
where $i =1$ and 2 distinguish between the two valence bonds in T1. 
Here, $\mu_i$ is the Lagrange multiplier used for imposing the hard-core constraint in the $i$-th bond.
Note that $H_{\textrm{T1},J}(s,t_\alpha)$ is the Heisenberg-model Hamiltonian
represented in terms of the $s$ and $t_\alpha$ bosons.
Explicitly, the above Hamiltonian can be written as follows:
\begin{eqnarray}
H_\textrm{T1} = \epsilon_{\textrm{T1},0} + H_{\textrm{T1},2} + H_{\textrm{T1},4},
 \label{eq:H^{NT}}
\end{eqnarray}
where
\begin{eqnarray}
\epsilon_{\textrm{T1},0}  = \sum_{i=1}^{2} \left[ - \frac{3}{4}J{\bar s}_i^{2} + \mu_i(1-{\bar s}_i^{2}) \right],
\end{eqnarray}
\begin{eqnarray}
H_{\textrm{T1},2} = \sum_{i=1}^{2} \left( \frac{J}{4} - \mu_i \right) t_{i\alpha}^{\dagger} t_{i\alpha},
\end{eqnarray}
\begin{eqnarray}
H_{\textrm{T1},4} = - \frac{\lambda J}{2} \epsilon_{\alpha\beta\gamma}\epsilon_{\alpha\mu\nu} t_{1\beta}^{\dagger}t_{1\gamma} t_{2\mu}^{\dagger}t_{2\nu}.
\label{eq:NT_quartic}
\end{eqnarray}
Note that all cubic terms containing three $t$ operators are ignored in the above since they play little role in a non-magnetic ground state.
It is interesting to observe that the quadratic Hamiltonian, $H_{\textrm{T1},2}$, is actually independent of the inter-dimer interaction strength, $\lambda J$.
This means that the ground state is a simple direct product of two spin singlet states. 
As mentioned previously, this is consistent with the exact result for the ground state.

The mean-field analysis is performed via quadratic decoupling of the quartic Hamiltonian, $H_{\textrm{T1},4}$.
With two mean-field parameters defined by $P=\langle t_{1\alpha} t_{2\alpha}^{\dagger} \rangle$ and $Q=\langle t_{1\alpha} t_{2\alpha} \rangle$,
the quadratically decoupled Hamiltonian of $H_{\textrm{T1},4}$ is given as follows:
\begin{eqnarray}
H^{\textrm{quad. dec.}}_{\textrm{T1},4}  &=& \frac{\lambda J}{3} ( Q^{*}Q - P^{*}P ) \nonumber\\
 &+& \frac{\lambda J}{3} \left( P t_{1\alpha}^{\dagger} t_{2\alpha} - Q t_{1\alpha}^{\dagger} t_{2\alpha}^{\dagger} + \textrm{H. c.} \right) . \label{eq:H_quad_dec}
\end{eqnarray}
For details of the derivation, see Appendix~\ref{appendix:Mean-field decoupling}.
Putting back Eq.~(\ref{eq:H_quad_dec}) into (\ref{eq:H^{NT}}),  we obtain the following mean-field Hamiltonian:
\begin{eqnarray}
 H^\textrm{MF}_\textrm{T1}
  &=&  \epsilon_{\textrm{T1},0} + \frac{\lambda J}{3} ( Q^{*}Q - P^{*}P )
  \nonumber\\
  &+& \frac{1}{2} {\bf t}_{\alpha}^{\dagger} {\bf M} {\bf t}_{\alpha} - \frac{3}{4} \textrm{Tr} {\bf M},
\end{eqnarray} 
where
\begin{eqnarray}
 {\bf M}
 = 
 \left(
 \begin{array}{cc|cc}
 \frac{J}{4}-\mu_1	& \frac{\lambda J}{3} P	& 0 	& -\frac{\lambda J}{3} Q
 \\
 \frac{\lambda J}{3} P^{*}	& \frac{J}{4}-\mu_2	& -\frac{\lambda J}{3} Q		& 0
 \\
 \hline
 0	& -\frac{\lambda J}{3} Q^{*} 	& \frac{J}{4}-\mu_1	& \frac{\lambda J}{3} P^{*}
 \\
 -\frac{\lambda J}{3} Q^{*}	& 0	& \frac{\lambda J}{3} P	&\frac{J}{4}-\mu_2
 \end{array}
 \right),
 \nonumber\\
\end{eqnarray}
and
\begin{eqnarray}
 {\bf t}_{\alpha}
 =
 \left(
 \begin{array}{c}
 t_{1\alpha}
 \\
 t_{2\alpha}
 \\
 \hline
 t_{1\alpha}^{\dagger}
 \\
 t_{2\alpha}^{\dagger}
 \end{array}
 \right).
\end{eqnarray}

The mean-field Hamiltonian, $H^\textrm{MF}_\textrm{T1}$, is diagonalized via the Bogoliubov transformation of ${\bf t}_{\alpha}$: 
\begin{eqnarray}
 H^\textrm{MF}_\textrm{T1}
 =
 \epsilon_{\textrm{gr},\textrm{T1}} +\sum_{i=1}^{2} \omega_i \gamma_{i\alpha}^{\dagger} \gamma_{i\alpha} ,
 \label{eq:H_{MF}^{NT}}
\end{eqnarray}
where 
\begin{eqnarray}
 \epsilon_{\textrm{gr},\textrm{T1}} = \epsilon_{\textrm{T1},0} +\frac{\lambda J}{3} ( Q^{*}Q - P^{*}P )  - \frac{3}{4} \textrm{Tr} {\bf M} + \frac{3}{2} \sum_{i=1}^{2} \omega_i
 \nonumber\\
\end{eqnarray}
and $\omega_i$ is the eigenfrequency of the $i$-th Bogoliubov quasiparticle, $\gamma_{i\alpha}$, which is in turn defined by
\begin{eqnarray}
 {\bf \Gamma}_{\alpha}
 =
 \left(
 \begin{array}{c}
  \gamma_{1\alpha}
  \\
  \gamma_{2\alpha}
  \\
  \hline
  \gamma_{1\alpha}^{\dagger}
  \\
  \gamma_{2\alpha}^{\dagger}
 \end{array}
 \right)
 =
 {\bf T} {\bf t}_{\alpha}
 \label{eq:bogoliubov_transf}
\end{eqnarray}
with ${\bf T}$ being the Bogoliubov transformation matrix. 
Note that the Bogoliubov transformation in the above is different from that of the BCS theory where a fermion (electron) is transformed to another fermion quasiparticle.  
Here, a boson, $t_\alpha$, is transformed to another boson quasiparticle, $\gamma_\alpha$.
To satisfy the canonical boson commutation relations, $\left[ {\bf \Gamma}_{\alpha}, {\bf \Gamma}_{\beta}^{\dagger} \right] = {\bf I}_B \delta_{\alpha {\beta}}$, where ${\bf I}_{B}$ is the diagonal matrix with elements being (1,1,-1,-1), 
the Bogoliubov transformation matrix, ${\bf T}$, is required to satisfy following constraint:
\begin{eqnarray}
 {\bf T} {\bf I}_{B} {\bf T}^{\dagger} = {\bf I}_{B} \;.
\end{eqnarray}
Due to this constraint, the diagonalization of the boson problem is slightly different from the fermion counterpart. The boson eigenvalue problem~\cite{BlaizotRipka} is given by 
\begin{eqnarray}
{\bf T} {\bf I}_{B} {\bf M} {\bf T}^{-1} = {\bf I}_{B} {\bf \Omega},
\end{eqnarray}
where ${\bf \Omega}$ is the diagonal matrix with elements being ($\omega_1$,$\omega_2$,$\omega_1$,$\omega_2$).

Assuming that every $\omega_i$ is non-negative for the stable valence bond ground state, $\left| \textrm{gr} \right>_{\rm{T1}}$, the ground state energy is equal to $\epsilon_{\textrm{gr},\textrm{T1}}$ which is a function of $\lambda$, ${\bar s}_i$, $\mu_i$, $P$, and $Q$. 
For a given $\lambda$, ${\bar s}_i$ and $\mu_i$ are self-consistently determined through the following four saddle point equations together with $P=\langle t_{1\alpha} t_{2\alpha}^{\dagger} \rangle$ and $Q=\langle t_{1\alpha} t_{2\alpha} \rangle$:
\begin{subequations}
\begin{align}
&\frac{\partial \epsilon_{\textrm{gr},\textrm{T1}}}{\partial {\bar s}_i} = \left( -\frac{3}{4}J - \mu_i \right) 2{\bar s}_i = 0,
\\
&\frac{\partial \epsilon_{\textrm{gr},\textrm{T1}}}{\partial \mu_i} = (1-{\bar s}_i^2) + \frac{3}{2} + \frac{\partial}{\partial \mu_i} \left( \frac{3}{2} \sum_{j=1}^{2} \omega_{j} \right) = 0,
\end{align}
\end{subequations}
where $i=1,2$.
The solution at $\lambda=0$ is simple and given by 
\begin{eqnarray}
{\bar s}_i = 1,~~~\mu_i = -\frac{3}{4}J,~~~P = 0,~~~Q = 0.
\end{eqnarray}
An important fact about the bond operator mean field theory is that the solution at $\lambda=0$ remains to be the solution at any $\lambda$ up to unity. 
This means that the ground state is simply given by the direct product of two spin singlet states:
\begin{eqnarray}
 \left| \textrm{gr} \right>_\textrm{T1} = 
 \left| s \right>_{ab} \otimes  \left| s \right>_{cd} ,
 \label{eq:NT_ground}
\end{eqnarray}
where $\left| s \right>_{ij}$ denotes the spin singlet wave function formed between the $i$ and $j$-th site.
Excited states are induced by breaking either the $(a,b)$ or the $(c,d)$ valence bond into a spin triplet, which results in a degenerate excitation energy of $1J$.

It is important to note that, in comparison with the exact states, the first excited state turns out to be grossly inaccurate while the ground state is correctly captured by the bond operator mean field theory. 
See  Table~\ref{tab:BOT_in_TO} for details of the result.
In the bond operator mean field theory, the wave function for the first excited state is given by  
\begin{equation}
 \gamma_{1\alpha}^{\dagger} \left| \textrm{gr} \right>_\textrm{T1} = 
 \left| t_{\alpha} \right>_{ab} \otimes  \left| s \right>_{cd} \;,
 \label{eq:first_exitation}
\end{equation}
while the exact solution at $\lambda=1$ is 
\begin{equation}
 \frac{1}{\sqrt{6}} \left( \left| t_{\alpha} \right>_{ab} \otimes \left| s \right>_{cd} +\left| t_{\alpha} \right>_{bd} \otimes \left| s \right>_{ca} + \left| t_{\alpha} \right>_{ad} \otimes \left| s \right>_{cb} \right) \;.
 \label{eq:exact_first_exitation}
\end{equation}
As one can see, similar to the bond operator result, the exact solution for the first excited state has a broken valence bond in the $(a,b)$ link.
However, due to additional spin correlations emerging from the $(a,d)$ as well as $(b,d)$ link, the final spin-spin correlation appears so that the spin singlet, which is supposed to form in the $(c,d)$ link, spreads into the $(a,c)$, $(b,c)$, and $(c,d)$ link equally with the spin-spin correlation value becoming $-0.417$. 
Consequently, the excitation energy of the exact first excited state, $0.5J$, is much lower than that of the bond operator mean field theory, $1J$.
Interestingly, the second excited state is well captured by the bond operator mean field theory.
To summarize, while correctly representing the ground (and the second excite) state, the bond operator theory description is rather inaccurate for the lowest spin excitation in the topologically orthogonal unit structure, T1.  
The conclusion so far is that the simple bond operator theory is not accurate enough to correctly capture the lowest spin excitation.
We believe that a promising direction for improvement is to construct a theory that incorporates the spin-spin correlations of the entire topologically orthogonal structure, not just between two nearest spins inside.
The spin cluster operator theory (SCOT) is such an attempt.

\begin{table}
 \centering
 \begin{ruledtabular}
 \begin{tabular}{cccccc}
  & & & \multicolumn{3}{c}{$\left< {\bf S}_i \cdot {\bf S}_j \right>$} \\
  State & Spin & Energy & $(a,b)$ & $(a,c),(b,c)$ & $(c,d)$\\
  \hline
  $
  \left| \textrm{gr} \right>_\textrm{T1}
  $
  &
  0
  &
  $
  \begin{array}{c}
  -1.5 J\\
  (-1.5 J)
  \end{array}
  $ 
  &
  $
  \begin{array}{c}
  -0.75 \\
  (-0.75)
  \end{array}
  $  
  &
  $
  \begin{array}{c}
  0 \\
  (0)
  \end{array}
  $ 
  &
  $
  \begin{array}{c}
  -0.75 \\
  (-0.75)
  \end{array}
  $  
  \\
  \hline
  $
  \gamma_{1\alpha}^{\dagger} \left| \textrm{gr} \right>_\textrm{T1}  
  $
  &
  1
  &
  $
  \begin{array}{c}
  -0.5 J\\
  (-1.0 J)
  \end{array}
  $ 
  &
  $
  \begin{array}{c}
  0.25 \\
  (0.25)
  \end{array}
  $  
  &
  $
  \begin{array}{c}
  0 \\
  (-0.417)
  \end{array}
  $ 
  &
  $
  \begin{array}{c}
  -0.75 \\
  (-0.417)
  \end{array}
  $ 
  \\
  \hline
  $
  \gamma_{2\alpha}^{\dagger} \left| \textrm{gr} \right>_\textrm{T1}  
  $
  &
  1
  &
  $
  \begin{array}{c}
  -0.5 J\\
  (-0.5 J)
  \end{array}
  $ 
  &
  $
  \begin{array}{c}
  -0.75 \\
  (-0.75)
  \end{array}
  $  
  &
  $
  \begin{array}{c}
  0 \\
  (0)
  \end{array}
  $ 
  &
  $
  \begin{array}{c}
  0.25 \\
  (0.25)
  \end{array}
  $ 
  \\
 \end{tabular}
 \end{ruledtabular}
 \caption{Various properties of the low energy states in the topologically orthogonal unit structure, T1, obtained from the bond operator mean field theory.  
 The exact diagonalization result is shown in parenthesis for comparison.
 \label{tab:BOT_in_TO}}
\end{table}

\begin{table}
 \centering
  \begin{ruledtabular}
  \begin{tabular}{ccccc}
  & & & \multicolumn{2}{c}{$\left< {\bf S}_i \cdot {\bf S}_j \right>$}\\
  State & Spin & Energy & \footnotesize{$(a,b),(c,d),(e,f)$}  & \footnotesize{$(b,d),(d,e),(e,b)$}\\
  \hline
  $
  \left| \textrm{gr} \right>_\textrm{E1}
  $
  &
  0
  &
  $
  \begin{array}{c}
  -2.524 J\\
  (-2.5 J)
  \end{array}
  $ 
  &
  $
  \begin{array}{c}
  -0.666 \\
  (-0.683)
  \end{array}
  $  
  &
  $
  \begin{array}{c}
  -0.176 \\
  (-0.15)
  \end{array}
  $ 
  \\
  \hline
  $
  \gamma_{m\alpha}^{\dagger}  \left| \textrm{gr} \right>_\textrm{E1}  
  $
  &
  1
  &
  $
  \begin{array}{c}
  -1.629 J\\
  (-1.764 J)
  \end{array}
  $ 
  &
  $
  \begin{array}{c}
  -0.572 \\
  (-0.370)
  \end{array}
  $  
  &
  $
  \begin{array}{c}
  0.029 \\
  (-0.218)
  \end{array}
  $ 
  \\
  \end{tabular}
  \end{ruledtabular}
 \caption{Various properties of the low energy states in the empty triangle unit structure, E1, obtained from the bond operator mean field theory.  
 The exact diagonalization results are shown in parenthesis for comparison.
 Note that the first excited states are sixfold degenerate and distinguished by $m=-1,1$ and $\alpha=x,y,z$.
 \label{tab:BOT_in_ET}}
\end{table}

Now, let us move to the empty triangle unit structure, E1, whose schematic diagram is shown in Fig.~\ref{fig:inter_dimer_int}. 
Relegating calculation details to Appendix~\ref{appendix:BOT_in_ET}, 
here, we discuss the results of the bond operator mean field theory
and compare them with exact diagonalization.
First of all, as one can see in Table~\ref{tab:BOT_in_ET},
the ground state energy as well as the spin-spin correlation values are in an excellent agreement with the exact diagonalization results.
This means that the ground state is quite well described by the bond operator mean field theory even though it is not the exact eigenstate.

As for the lowest excitation, the bond operator theory predicts sixfold degenerate spin excitations, which is consistent with the exact results.
Moreover, while there are some discrepancies between the spin-spin correlation value of the bond operator mean field theory and exact diagonalization,  
the energy itself is very well reproduced by the bond operator mean field theory.  
Considering the excellent agreement in energy, we believe that the bond operator mean field theory works reasonably well in the case of the empty triangle structure. 
It is important to note that the inter-dimer interaction is completely ignored by the bond operator mean field theory in the topologically orthogonal structure while it is not true in the empty triangle counterpart.

The final conclusion of this section is that the ground state is quite well captured by the bond operator mean field theory in the both topologically orthogonal and empty triangle structure. 
The lowest energy excitation is, however, poorly represented in the topologically orthogonal structure since the simple bond operator theory regards a pair of the nearest neighboring spins as the basic building block. 
We believe that it is necessary to extend the basic building block of the system dynamics from a dimer to a spin cluster to incorporate as many topologically orthogonal structures as possible.

\subsection{Binding dimers into a spin cluster}

As mentioned in the preceding section, for the correct representation of low-energy spin excitations, it is necessary to bind dimers into a spin cluster so that the spin-spin correlation in the topologically orthogonal structure is better taken into account. 
To distinguish from the conventional bond operator theory, we call such theory the spin cluster operator theory (SCOT).

As depicted in Fig.~\ref{fig:Cluster covering}, there are two possible patterns for the VBS order with a 36-site unit cell: 
the VBS order with (i) staggered hexagonal resonance and (ii) uniform hexagonal resonance. 
Appropriate spin clusters are selected according to the following conditions:
(i) the VBS order formed inside each spin cluster should be consistent with that of the whole lattice, 
(ii) the shape of the spin clusters should preserve the symmetry of the VBS order, and
(iii) the spin clusters should include as many topologically orthogonal structures as possible within themselves. 
The most general spin cluster that satisfies all these three conditions is in fact the unit cell itself, where every inter-dimer interaction is topologically orthogonal. 
Dubbed as the ``super cluster,'' however, the whole unit cell is too large to be diagonalized directly since a very large number of spin triplet excited states are necessary to correctly represent the physical spin.
In this study, we choose a particular subset of the unit cell that is consistent with all the above three conditions.

Figure~\ref{fig:Cluster covering} shows the spin cluster covering for the VBS order with  staggered and uniform hexagonal resonance.
Let us first consider the VBS order with staggered hexagonal resonance, in which case the unit cell is composed of three spin clusters: a David star, D, and two extended hexagons, H1 and H2.
The David-star spin cluster consists of six dimers which form a pinwheel structure.
Extended hexagons are created by binding the rest of the dimers in such a way that the resulting spin cluster covering preserves the sixfold (threefold) rotation symmetry around the center of the pinwheel (perfect hexagon) structure.
It can be shown that the above separation of the unit cell into subset spin clusters is fully consistent with the VBS order of the whole lattice.
For the VBS order with uniform hexagonal resonance, there are two different kinds of the spin clusters in the unit cell: (i) a core cluster, C, composed of a David star and three attached dimers and (ii) nine dimers, S1 - S9.
It is confirmed again that such separation into subset spin clusters is consistent with the VBS order with uniform hexagonal resonance, which has a threefold rotation symmetry around the center of the pinwheel structure and a reflection symmetry with respect to the vertical axis piercing through the pinwheel structure.
Now that appropriate spin clusters are selected for each VBS order,
we develop a fully fledged spin cluster operator theory in the next section.

\section{SPIN CLUSTER OPERATOR THEORY: STAGGERED HEXAGONAL RESONANCE 
\label{sec:SCOT_I}}

\subsection{Spin cluster operator representation}

\begin{figure}
 \begin{minipage}{0.3\linewidth}
  \centering%
  \includegraphics[width=1.0\textwidth]{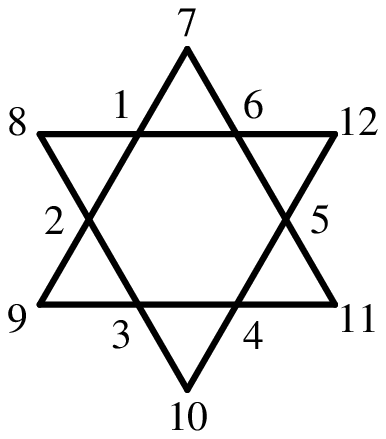}
 \end{minipage}
 \begin{minipage}{0.5\linewidth}
  \centering%
  \includegraphics[width=1.0\textwidth]{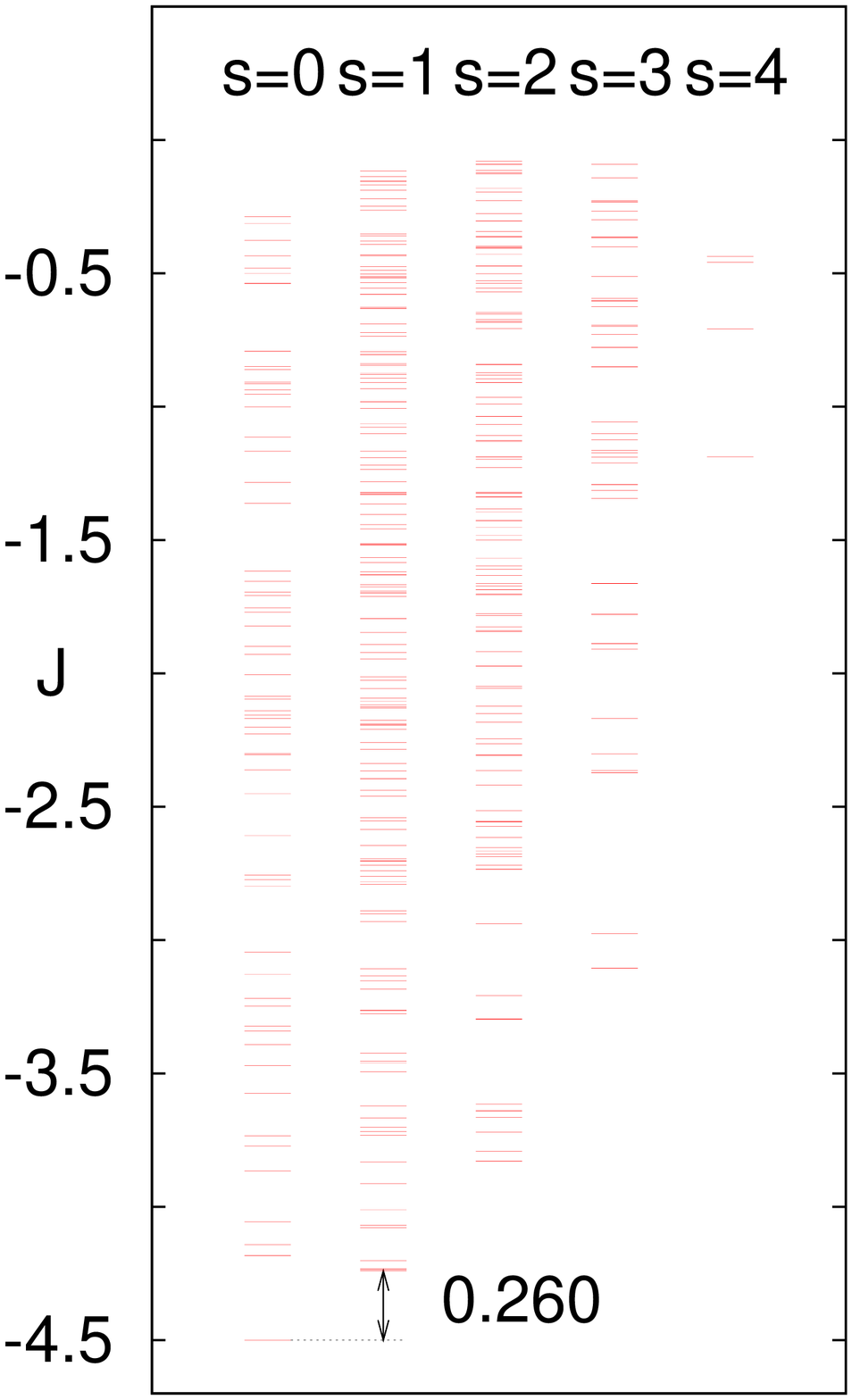}
 \end{minipage}
 \\
 \begin{minipage}[b]{\linewidth}
  \centering%
  \caption{(Color online)
  Schematic diagram for the David-star spin cluster and the excitation energy spectrum showing two thousands states above two degenerate ground states at $S=0$.
  \label{fig:cluster_C}}
 \end{minipage}
\end{figure}

\subsubsection{David star}

Constructing the spin cluster operator representation is actually nothing but choosing an appropriate basis set suitable for representing the low energy states in the presence of the VBS order.  
In choosing a suitable basis set for the spin cluster, first noted is the fact that the number of available states increases exponentially as the size of the spin cluster grows. 
For example, in a David star spin cluster, there are $2^{12}$ states with various spin values ranging from $S=0$ to 6.   
The figure~\ref{fig:cluster_C} shows two thousand low energy states emerging from a David star spin cluster.
As a matter of principle, all eigenstates should be retained for the exact representation of the physical spin.
Setting aside the technical issue that the number of eigenstates is too large, an important physical question is how to deal with the large spin states.
In this work, we take an approach that, first, the lowest energy state in the $S=0$ channel is condensed so that the appropriate VBS order is formed in the ground state.
Second, with no clear energy gap in the $S=1$ channel dividing between the low and high region of the spectrum, we include as many spin triplet states as possible for the representation of the physical spin. 
Technically, our approach is to increase the number of spin triplet states until results such as the ground state energy as well as the spin gap converge into a well-defined fixed value.  
Finally, assuming the the lowest spin excitations are spin triplet, i.~e., in the $S=1$ channel, we ignore all other higher spin states for the spin cluster representation.

As mentioned in the preceding section, the ground state in the spin singlet channel is degenerate with two different orientations for the dimer covering pattern in the pinwheel structure.
Mathematically, the ground-state Hilbert space is spanned by the following two VBS states:
\begin{subequations}
\label{eq:phi}
\begin{eqnarray}
 &&\left| \phi_1 \right> = \bigotimes_{(i,j)\in I_1}  \left| s \right>_{ij},
 \label{eq:phi_1}
\end{eqnarray}
\begin{eqnarray}
 &&\left| \phi_2 \right> = \bigotimes_{(i,j)\in I_2}  \left| s \right>_{ij},
  \label{eq:phi_2}
\end{eqnarray}
\end{subequations}
where $I_1$=$\{$(1,8),(2,9),(3,10),(4,11),(5,12),(6,7)$\}$ and $I_2$=$\{$(1,7),(2,8),(3,9),(4,10),(5,11),(6,12)$\}$. 
The above two states are not orthogonal as seen from the non-zero overlap: $\left< \phi_1 | \phi_2 \right>=1/32$.
While it is straightforward to construct an orthonormal basis set out of the above two states:
\begin{subequations}
\label{eq:psi}
\begin{eqnarray}
 \left| \psi_1 \right> = \frac{1}{\sqrt{2+1/16}} \left( \left| \phi_1 \right> + \left| \phi_2 \right> \right),
 \label{eq:psi_1}
\end{eqnarray}
\begin{eqnarray}
 \left| \psi_2 \right> = \frac{1}{\sqrt{2-1/16}} \left( \left| \phi_1 \right> - \left| \phi_2 \right> \right),
 \label{eq:psi_2}
\end{eqnarray}
\end{subequations}
the new states, $\left| \psi_1 \right>$ and $\left| \psi_2 \right>$, no longer possess the VBS order.
In light of our assumption that the VBS order exists in the ground state, we take either $\left| \phi_1 \right>$ or $\left| \phi_2 \right>$ to be condensed.  
It does not matter which one of the two states is condensed since the reflection symmetry connecting between $\left| \phi_1 \right>$ and $\left| \phi_2 \right>$ is presumed to be broken spontaneously.

As for the excited states in the $S=1$ channel, basis states are chosen first as the eigenstates of ${\bf S}^2$ (${\bf S}=\sum_{i=1}^{12}{\bf S}_i$ and S$_\alpha$ ($\alpha=x,y,z$) in the following combination:
\begin{subequations}
\label{eq:triplet basis}
\begin{align}
 &\left| t_x \right> = \frac{-1}{\sqrt{2}} \left( \left| S=1,m=1 \right> - \left| S=1,m=-1 \right> \right),
 \label{eq:triplet basis x}
 \\
 &\left| t_y \right> = \frac{i}{\sqrt{2}} \left( \left| S=1,m=1 \right> + \left| S=1,m=-1 \right> \right),
 \label{eq:triplet basis y}
 \\
 &\left| t_z \right> = \left| S=1,m=0 \right>.
 \label{eq:triplet basis z}
\end{align}
\end{subequations}
Note that each $\left| t_\alpha \right>$ is invariant under their respective spin rotation, i.~e.,
\begin{eqnarray}
 \textrm{S}_x \left| t_x \right> = \textrm{S}_y \left| t_y \right> = \textrm{S}_z \left| t_z \right> = 0.
 \label{eq:triplet_basis_property}
\end{eqnarray} 
Unfortunately, the above classification does not uniquely choose the basis states since there exist degeneracies among many of the spin triplet states.
This problem can be resolved by considering the sixfold space rotation operator, R$_6$, which is defined as follows:
\begin{align}
 &\textrm{R}_6 \left| m_1,m_2,m_3,m_4,m_5,m_6,m_7,m_8,m_9,m_{10},m_{11},m_{12} \right>	\nonumber\\
 &= \left| m_6,m_1,m_2,m_3,m_4,m_5,m_{12},m_7,m_8,m_9,m_{10},m_{11} \right>,
\end{align}
where $m_i$ denotes the eigenvalue of S$_{i,z}$.
The eigenvalue of R$_6$ is given by an element of the set, $\mathbb{R}_6 \equiv \{ e^{i 2 \pi l / 6} | l=-2,-1,0,1,2,3 \}$.
Degenerate spin triplet states can be decomposed into the eigenstates of R$_6$ since it commutes with both the Hamiltonian, $J \sum_{\left< i , j \right>} {\bf S}_i \cdot {\bf S}_j$, and the square of the total spin operator, ${\bf S}^2$.
Degeneracy in energy can appear between $l$'s with different signs, but the same magnitude, i.~e., either $l=\pm 1$ or $l=\pm 2$. 

Taking the eigenstates of R$_6$ provides an additional advantage for the  construction of the spin cluster representation. 
If it were not for R$_6$, the spin cluster representation has to be computed for every individual spin. 
With help of the sixfold space rotation symmetry, all spin operators can be computed from the knowledge of only two spin operators, ${\bf S}_1$ and ${\bf S}_7$:
\begin{subequations}
\begin{eqnarray}
 {\bf S}_{1+p} = \textrm{R}_6^{p} {\bf S}_{1} \textrm{R}_6^{-p},
\end{eqnarray}
\begin{eqnarray}
 {\bf S}_{7+p} = \textrm{R}_6^{p} {\bf S}_{7} \textrm{R}_6^{-p}
\end{eqnarray}
\end{subequations}
for $p=0, \dots, 5$.

The spin cluster operator representation of the physical spin is formally written as follows:
\begin{eqnarray}
 \textrm{S}_{i+p,\alpha}
  &=& \sum_{\mu,\nu} \hat{O}_\mu^{\dagger} \left< \mu \right| \mathrm{R}_6^{p} \textrm{S}_{i\alpha} \mathrm{R}_6^{-p}\left| \nu \right> \hat{O}_\nu,
\end{eqnarray}
where $i=1,7$ and $p=0,\cdots,5$.
Here, $\left| \mu \right>$ is the $\mu$-th energy eigenstate and $\hat{O}_\mu^{\dagger}$ denotes the corresponding creation operator.
Restricted to the Hilbert space with the ground state (which has a VBS order) and the spin triplet excitations, the spin cluster operator representation of the physical spin is finally obtained as follows:
\begin{align}
  &\textrm{S}_{i+p,\alpha}
  = \sum_{n=1}^{M} \left[ \left( f_{n}^{(i)} \cdot z_n^{p} \right) t_{n\alpha}^{\dagger} s + \textrm{H. c.} \right]
  \nonumber\\
  &+ \sum_{m,n=1}^{M} \left( e_{mn}^{(i)} \cdot z_m^{p} z_n^{-p} \right) \epsilon_{\alpha\beta\gamma} t_{m\beta}^{\dagger} t_{n\gamma} 
  \label{eq:David_representation}
\end{align}
where $s^\dagger$ is the creation operator for the spin singlet ground state with one of the two possible pinwheel configurations, $\left| \phi_1 \right>$ and $\left| \phi_2 \right>$, in Eq.~(\ref{eq:phi}).
$t^\dagger_{n\alpha}$ is the creation operator for the spin triplet state which is a simultaneous eigenstate of the Hamiltonian, $H$, the sixfold space rotation operator, R$_6$, and the spin operator, S$_\alpha$:
\begin{subequations}
\begin{align}
H \left| t_{n\alpha} \right> &= \epsilon_{n}^{D} \left| t_{n\alpha} \right> ,\\
\textrm{R}_6 \left| t_{n\alpha} \right> &= z_n \left| t_{n\alpha} \right> ~ (z_n \in \mathbb{R}_6), \\
\textrm{S}_\alpha \left| t_{n\alpha} \right> &= 0 ,
\end{align}
\end{subequations}
where $\epsilon_{n}^{D}$ is the eigenenergy of the $n$-th spin triplet excitation in the David star and $z_n$ is the eigenvalue of R$_6$ belonging to $\mathbb{R}_6$.
In the above, $M$ is the total number of spin triplet states included in spin cluster operator representation.
The matrix elements, $f_{n}^{(i)}$ and $e_{mn}^{(i)}$, are defined by
\begin{align}
f_{n}^{(i)} &= \left< t_{nx} \right| \textrm{S}_{ix} \left| s \right>, 
\label{eq:f_n}\\
e_{mn}^{(i)} &= \left< t_{my} \right| \textrm{S}_{ix} \left| t_{nz} \right>,
\label{eq:e_mn}
\end{align}
where it is interesting to note that $e_{mn}^{(i)}$ is anti-hermitian.
A brief sketch for the derivation of the spin cluster operator representation is given in Appendix~\ref{appendix:SCOT_representation}.

\subsubsection{Extended hexagon}

\begin{figure}
 \begin{minipage}{0.4\linewidth}
  \centering%
  \includegraphics[width=1.0\textwidth]{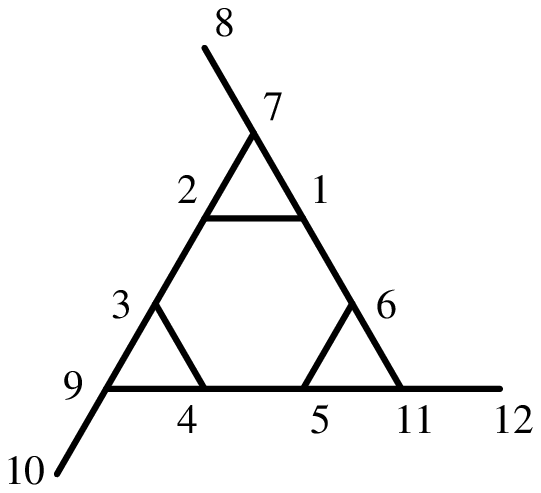}
 \end{minipage}
 \begin{minipage}{0.5\linewidth}
  \centering%
  \includegraphics[width=1.0\textwidth]{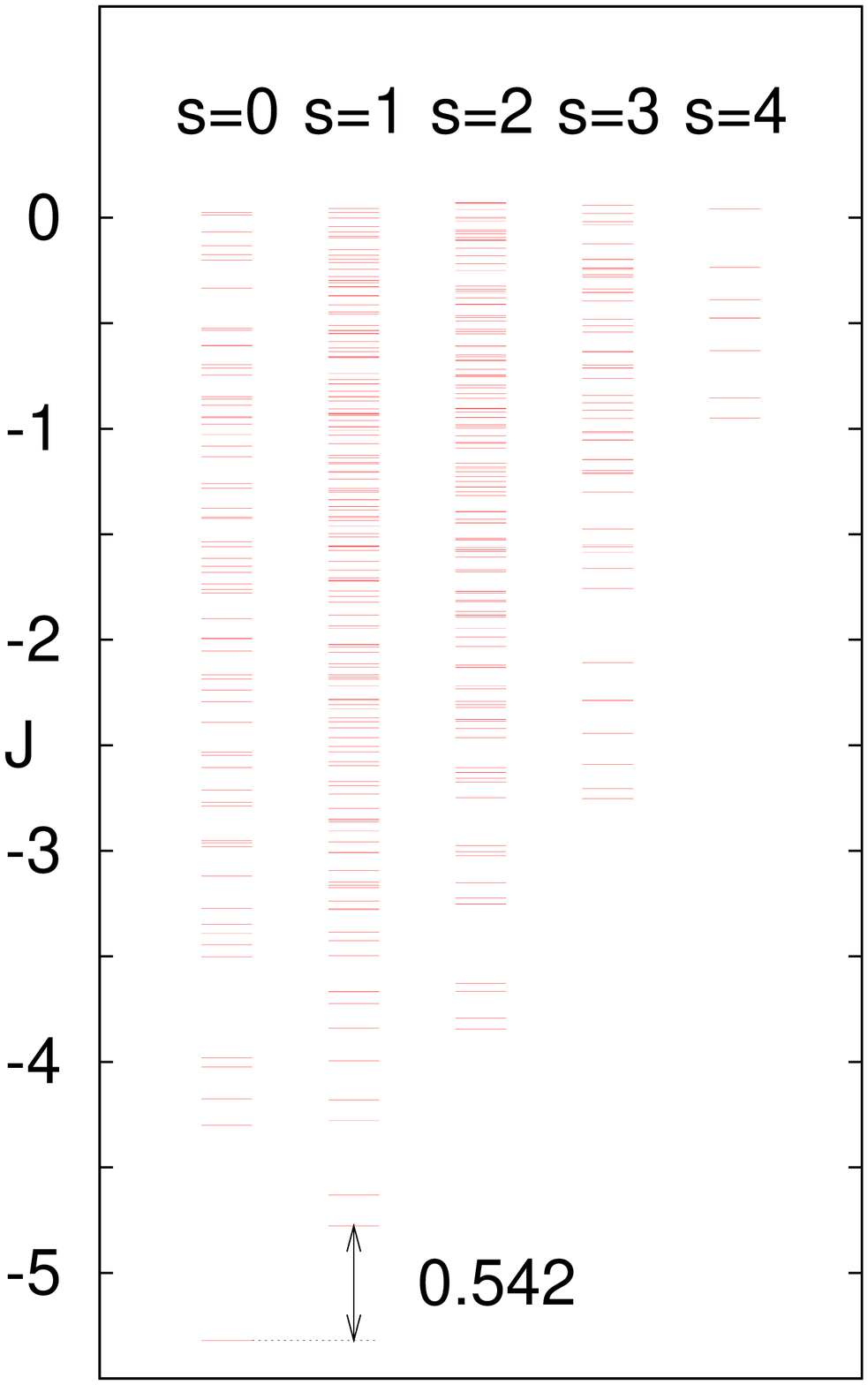}
 \end{minipage}
 \\
 \begin{minipage}[b]{\linewidth}
  \centering%
  \caption{(Color online)
  Schematic diagram for the extended hexagon spin cluster and the excitation energy spectrum showing two thousands states above the spin-singlet ground state.
  \label{fig:cluster_H}}
 \end{minipage}
\end{figure}

As mentioned previously, the unit cell of the VBS ground state with staggered hexagonal resonance is composed of a David star and two extended hexagons with three attached dimers.
The spin cluster operator representation for the extended hexagon is constructed in the similar way to the David star case with two minor modifications.  
First, the ground state has no degeneracy in the extended hexagon spin cluster.
Second, the extended hexagon spin cluster has a threefold space rotation symmetry. 
With the threefold space rotation operator, R$_3$, defined by
\begin{align}
 &\textrm{R}_3 \left| m_1,m_2,m_3,m_4,m_5,m_6,m_7,m_8,m_9,m_{10},m_{11},m_{12} \right>	   
 \nonumber\\
 &= \left| m_5,m_6,m_1,m_2,m_3,m_4,m_{11},m_{12},m_7,m_8,m_9,m_{10} \right>,
 \nonumber\\
\end{align}
the eigenvalue of R$_3$ is an element of the set, $\mathbb{R}_3 \equiv \{ e^{i 2 \pi l / 3} | l=-1,0,1 \}$.
As before, the spin cluster operator representation of the four spin operators, ${\bf S}_1$, ${\bf S}_2$, ${\bf S}_7$, and ${\bf S}_8$, determine those of other eight spins with help of the threefold rotation symmetry:
\begin{eqnarray}
 {\bf S}_{i+2p} = \textrm{R}_{3}^{p} {\bf S}_{i} \textrm{R}_{3}^{-p},
\end{eqnarray}
where $i=1,2,7,8$ and $p=1,2$.

Repeating essentially the same procedure described in the previous section, the spin cluster operator representation for the extended hexagon is obtained as follows:
\begin{align}
 &\textrm{S}_{i+2p,\alpha}
  =\sum_{n=1}^{N} \left[ ( a_{n}^{(i)} \cdot r_n^{p} ) t_{n\alpha}^{\dagger} s + \textrm{H. c.} \right]
  \nonumber\\
  &+\sum_{m,n=1}^{N} \left( b_{mn}^{(i)} \cdot r_m^{p} r_n^{-p} \right) \epsilon_{\alpha\beta\gamma} t_{m\beta}^{\dagger} t_{n\gamma},
\end{align}
where
\begin{align}
 \mathrm{R_3} \left| s \right> &= \left| s \right>, \\
 \mathrm{R_3} \left| t_{n\alpha} \right> &= r_n \left| t_{n\alpha} \right> ~ (r_n \in \mathbb{R}_3), \\
 a_{n}^{(i)} &= \left< t_{nx} \right| \textrm{S}_{ix} \left| s \right>, 
 \label{eq:a_n}\\
 b_{mn}^{(i)} &= \left< t_{my} \right| \textrm{S}_{ix} \left| t_{nz} \right>,
 \label{eq:b_mn}
\end{align}
and $N$ is the total number of spin triplet states considered for the extended hexagon spin cluster.

\subsection{Hamiltonian and the mean field theory}

The basic plan for analysis is to rewrite the Heisenberg-model Hamiltonian in the spin cluster operator representation, which is obtained in the preceding section, and apply the self-consistent mean-field method similar to the bond operator theory as described in Section~\ref{subsec:testing_BOT}. 
Let us begin by formally writing the Heisenberg-model Hamiltonian in the spin cluster operator representation, accompanied by three Lagrange multiplier terms each corresponding to the hard-core constraint in a David star and two extended hexagons:
\begin{align}
 H=H_J
 &-\mu_{D} \sum_{\bf r} \sum_{n=1}^{M} \left[ {\bar s}_{D}^2 +  t^{D\dagger}_{n\alpha}({\bf r}) t^{D}_{n\alpha}({\bf r}) - 1 \right] \nonumber\\
 &-\mu_{H} \sum_{\bf r} \sum_{n=1}^{N} \left[ {\bar s}_{H}^2 +  t^{H1\dagger}_{n\alpha}({\bf r}) t^{H1}_{n\alpha}({\bf r}) - 1 \right]  \nonumber\\
 &-\mu_{H} \sum_{\bf r} \sum_{n=1}^{N} \left[ {\bar s}_{H}^2 +  t^{H2\dagger}_{n\alpha}({\bf r}) t^{H2}_{n\alpha}({\bf r}) - 1 \right], 
\end{align}  
where
\begin{align}
 H_J = J \sum_{\left< i,j \right> \in {\cal C}} {\bf S}_i \cdot {\bf S}_j
 + \lambda J \sum_{\left< i,j \right> \notin {\cal C}} {\bf S}_i \cdot {\bf S}_j,
\end{align}
with ${\cal C}$ denoting the site-index pairs belonging to the
valence bond in the David star, $D$, as well as the two extended hexagon spin clusters, $H1$ and $H2$.
As before, $\lambda$ is introduced for computational convenience and is set to be unity at the end of computation. 
${\bar s}$  and $\mu$ denotes the spin singlet condensation density and the Lagrange multiplier for the hard-core constraint, respectively.

Arranging the Hamiltonian with respect to the degree of spin triplet operators leads to
\begin{eqnarray}
 H = \mathscr{N} \epsilon_0 + H_2 + H_4,
\end{eqnarray}
where $\mathscr{N}$ is the total number of unit cells.
$H_2$ and $H_4$ is the quadratic and quartic part of the Hamiltonian, respectively.
The constant term in the Hamiltonian, $\epsilon_0$, is given by
\begin{align}
 \epsilon_0
 &=\epsilon^{D}_{0} {\bar s}_{D}^{2} + \mu_{D} (1-{\bar s}_{D}^{2})	\nonumber\\
 &+\epsilon^{H}_{0} {\bar s}_{H}^{2} + \mu_{H} (1-{\bar s}_{H}^{2})	\nonumber\\
 &+ \epsilon^{H}_{0} {\bar s}_{H}^{2} + \mu_{H} (1-{\bar s}_{H}^{2}) ,
 \label{eq:SCOT_e_o}
\end{align}
where $\epsilon^{D}_0$ and $\epsilon^{H}_0$ is the ground state energy of the David star and the extended hexagon spin clusters, respectively.
The mean field Hamiltonian is obtained via self-consistent quadratic decoupling of the quartic terms  
as described in Appendix~\ref{appendix:Mean-field decoupling}.
Mean-field parameters are defined as follows:
\begin{subequations}
\begin{eqnarray} 
 \mathscr{P}_{kl}^{DH1}({\bf r'})
 &=& \left< t_{k\alpha}^{D}({\bf r}) t_{l\alpha}^{H1\dagger}({\bf r}+{\bf r'}) \right>,\\
 \mathscr{P}_{kl}^{DH2}({\bf r'})
 &=& \left< t_{k\alpha}^{D}({\bf r}) t_{l\alpha}^{H2\dagger}({\bf r}+{\bf r'}) \right>,\\
 \mathscr{P}_{kl}^{HH}({\bf r'})
 &=& \left< t_{k\alpha}^{H1}({\bf r}) t_{l\alpha}^{H2\dagger}({\bf r}+{\bf r'}) \right>,
\end{eqnarray}
\label{eq:SCOT_P}
\end{subequations}
and
\begin{subequations}
\begin{eqnarray}
 \mathscr{Q}_{kl}^{DH1}({\bf r'})
 &=& \left< t_{k\alpha}^{D}({\bf r}) t_{l\alpha}^{H1}({\bf r}+{\bf r'}) \right>, \\
 \mathscr{Q}_{kl}^{DH2}({\bf r'})
 &=& \left< t_{k\alpha}^{D}({\bf r}) t_{l\alpha}^{H2}({\bf r}+{\bf r'}) \right>, \\
 \mathscr{Q}_{kl}^{HH}({\bf r'})
 &=& \left< t_{k\alpha}^{H1}({\bf r}) t_{l\alpha}^{H2}({\bf r}+{\bf r'}) \right>,
\end{eqnarray}
\label{eq:SCOT_Q}
\end{subequations}
with ${\bf r'}$ being one of the following vectors:
\begin{subequations}
\begin{eqnarray}
 {\bf r}_A &=& 4\sqrt{3} a \left( -\frac{\sqrt{3}}{2} \hat{x} -\frac{1}{2} \hat{y} \right),
 \\
 {\bf r}_B &=& 4\sqrt{3} a \left( -\hat{y} \right),
 \\
 {\bf r}_C &=& {\bf r}_B - {\bf r}_A ,
\end{eqnarray}
\end{subequations}
where $a$ is the distance between the nearest neighboring spins.

The mean field Hamiltonian can be written in a compact form via the Fourier transformation:
\begin{align}
 H_\textrm{MF}
 &= \mathscr{N} \epsilon_0 + \mathscr{N} \epsilon_{\mathscr{PQ}} 
 \nonumber\\
 &+ \frac{1}{2} \sum_{{\bf k}} \boldsymbol{\tau}_{\alpha}^{\dagger}({\bf k}) {\bf M}({\bf k}) \boldsymbol{\tau}_{\alpha}({\bf k}) - \frac{3}{4} \sum_{\bf k} \textrm{Tr}{\bf M}({\bf k}),
 \nonumber\\
 \label{eq:H_{MF}}
\end{align}
where
\begin{align}
 \boldsymbol{\tau}_{\alpha}({\bf k})
  =
 \left(
  \begin{array}{c}
   {\bf t}_{\alpha}^{D}({\bf k})	\\
   {\bf t}_{\alpha}^{H1}({\bf k})	\\
   {\bf t}_{\alpha}^{H2}({\bf k})	\\
   \hline
   {\bf t}_{\alpha}^{D\dagger}(-{\bf k})	\\
   {\bf t}_{\alpha}^{H1\dagger}(-{\bf k})	\\
   {\bf t}_{\alpha}^{H2\dagger}(-{\bf k})	\\
   \end{array}
 \right) .
\end{align}
In the above, ${\bf t}^\dagger_\alpha({\bf k})$ and ${\bf t}_\alpha({\bf k})$ are the respective column vectors composed of the creation and annihilation operators for spin triplet states.
The size of the $\boldsymbol{\tau}$ vector is given by $2(M+2N)$.
The detailed form of $\epsilon_{\mathscr{PQ}}$ and ${\bf M}({\bf k})$ are given in 
Appendix~\ref{appendix:details_of_the_mean_field_Hamiltonian}.

Diagonalizing the mean field Hamiltonian via the Bogoliubov transformation leads to the following ground state energy per spin:
\begin{align}
 \epsilon_{gr} &=\frac{1}{36} \left[ \right. \epsilon_0 + \epsilon_{\mathscr{PQ}} \nonumber\\
 & -\frac{3}{4} \frac{1}{\mathscr{N}}\sum_{\bf k}\textrm{Tr}{\bf M}({\bf k}) + \frac{3}{2} \frac{1}{\mathscr{N}}\sum_{\bf k}\sum_{n=1}^{M+2N} \omega_{n}({\bf k}) \left. \right] ,
 \nonumber\\
\end{align}
where $\omega_{n}({\bf k})$ is the solution of the eigenvalue problem:
\begin{equation}
 {\bf T}({\bf k}) {\bf I}_B {\bf M}({\bf k}) {\bf T}^{-1}({\bf k}) =  {\bf I}_B \boldsymbol{\Omega}({\bf k}),
\label{inf_eigenvalue_problem}
\end{equation}
where
\begin{align}
 {\bf I}_{B}
 &=
 \left(
 \begin{array}{cc}
 {\bf I} & {\bf 0}
  \\
 {\bf 0} & -{\bf I}
 \end{array}
 \right),
 \\
 \boldsymbol{\Omega}({\bf k})
 &=
 \left(
 \begin{array}{cc}
 \boldsymbol{\omega}({\bf k}) & {\bf 0}
  \\
 {\bf 0} & \boldsymbol{\omega}({\bf k})
 \end{array}
 \right).
\end{align}
As before, the spin singlet condensation density as well as the Lagrange multiplier for the hard-core constraint are determined by the saddle-point equations:
\begin{eqnarray}
 \frac{\partial \epsilon_{gr}}{\partial {\bar s}} = \frac{\partial \epsilon_{gr}}{\partial \mu} = 0,
 \label{SCF}
\end{eqnarray}
where ${\bar s}={\bar s}_D$ or ${\bar s}_{H}$, and $\mu=\mu_D$ or $\mu_{H}$.
The saddle-point equations are solved simultaneously under the self-consistent conditions for the mean field parameters defined in Eq.~(\ref{eq:SCOT_P}) and (\ref{eq:SCOT_Q}).
With a solution of the full parameter set, (${\bar s}$, $\mu$, $\mathscr{P}$, $\mathscr{Q}$), simultaneously satisfying all the saddle-point and the self-consistence equations at $\lambda=1$, 
one can compute the ground state energy as well as the spin triplet excitation spectrum.
Results are discussed in Section~\ref{sec:RESULTS_I}.
\\
\\

\section{SPIN CLUSTER OPERATOR THEORY: UNIFORM HEXAGONAL RESONANCE
\label{sec:SCOT_II}}

\subsection{Spin cluster operator representation}

\begin{figure}
 \centering
 \includegraphics[width=0.5\linewidth]{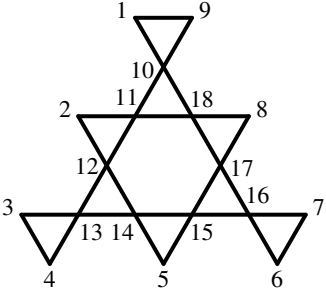}
 \caption{Schematic diagram for the core cluster in the VBS order with uniform hexagonal resonance.
 \label{fig:core_cluster}}
\end{figure}

The core cluster shown in Fig.~\ref{fig:core_cluster} has a twofold degeneracy in the ground state energy, which is related to two possible dimer orientations in the pinwheel structure.
That is to say, either of the following two spin singlet states can be condensed for the ground state:
\begin{subequations}
\begin{eqnarray}
 \left| \pi_1 \right> = \bigotimes_{(i,j)\in I \cup {J}_1} \left| s \right>_{ij},
 \label{eq:pi_1}
\end{eqnarray}
\begin{eqnarray}
 \left| \pi_2 \right> = \bigotimes_{(i,j)\in I \cup {J}_2}  \left| s \right>_{ij},
 \label{eq:pi_2}
\end{eqnarray}
\end{subequations}
where I=$\{$(9,1), (3,4), (6,7)$\}$, J$_1$=$\{$(11,2), (12,13), (14,5), (15,16), (17,8), (18,10)$\}$, and J$_2$=$\{$(10,11), (2,12), (13,14), (5,15), (16,17), (8,18)$\}$.

By using the similar method for the analysis of the David star as well as the extended hexagon cluster, the spin cluster operator representation for the core cluster is obtained as follows:
\begin{align}
 &\textrm{S}_{i+3p,\alpha}
  =\sum_{n=1}^{L} \left[ ( \tilde{a}_{n}^{(i)} \cdot \tilde{r}_n^{p} ) t_{n\alpha}^{\dagger} s +\textrm{H. c.} \right]
  \nonumber\\
  &+\sum_{m,n=1}^{L} ( \tilde{b}_{mn}^{(i)} \cdot \tilde{r}_m^{p} \tilde{r}_n^{-p} ) \epsilon_{\alpha\beta\gamma} t_{m\beta}^{\dagger} t_{n\gamma},
  \label{core_rep}
\end{align}
where $i=1,2,3$ and $p=0,1,2$. 
Also, in the above,
\begin{align}
 \left| s \right> &\in \{ \left| \pi_1 \right>, \left| \pi_2 \right> \}, \\
 \mathrm{R_3} \left| s \right> &= \left| s \right>, \\
 \mathrm{R_3} \left| t_{n\alpha} \right> &= \tilde{r}_n \left| t_{n\alpha} \right> (\tilde{r}_n \in \mathbb{R}_3), \\
 \tilde{a}_{n}^{(i)} &= \left< t_{nx} \right| \textrm{S}_{ix} \left| s \right>, \\
 \tilde{b}_{mn}^{(i)} &= \left< t_{my} \right| \textrm{S}_{ix} \left| t_{nz} \right>,
\end{align}
and $L$ is the total number of spin triplet states included for the core cluster representation.

In addition to the core cluster, there remain nine dimers surrounding the core.  
See Fig.~\ref{fig:Cluster covering} for the illustration of the VBS unit cell with uniform hexagonal resonance.
The nine surrounding dimers can be represented by the usual bond operator formalism discussed in Sec.~\ref{subsec:testing_BOT}.

\subsection{Hamiltonian and the mean field theory}

The analysis of the mean field Hamiltonian for the VBS order with uniform hexagonal resonance is similar to that of the staggered hexagonal resonance counterpart.
Since computational details are basically identical, we skip the technical discussion and just mention a few points that are specific to the VBS order with uniform hexagonal resonance. 
First, once the core cluster is isolated from the unit cell, there is no natural spin cluster with an extended size compatible with the VBS order.  
For this reason, all remaining spins are paired as dimers and analyzed via the bond operator theory. 
Second, it turns out that the core cluster is completely decoupled from the surrounding dimers so that spin triplet excitations are separated into two independent channels with one being confined into the core cluster and the other propagating through the surrounding dimers. 
Details of the results are given in Sec.~\ref{sec:RESULTS_II}.

\section{RESULTS: STAGGERED HEXAGONAL RESONANCE \label{sec:RESULTS_I}}

The results of the spin cluster operator theory for the VBS order with staggered hexagonal resonance are discussed in this section.
In addition to the thermodynamic limit where the site number goes to infinity,
we also report the results for the 36-site unit cell, which can be directly compared with the exact diagonalization results obtained in the same-sized system.
Here, the 36-site unit cell means that the periodic boundary condition is used without the appropriate phase factors required for the infinite lattice.

In the both cases of the infinite lattice and the 36-site unit cell, computations are performed as a function of the number of spin triplet states included in the spin cluster operator representation. 
With the number of included spin triplet states denoted as $M$ for the David star and $N$ for each extended hexagon cluster, the total spin-triplet state number is increased along $M=N$ in this study.
It is found in the end of computation that the ground state energy as well as the lowest spin excitation gap are quite well converged already for $M \sim 80$.
Note that various parallel computation techniques are used in this study to cope with the complexity associated with a large number of self-consistent mean field parameters defined in Eq.~(\ref{eq:SCOT_P}) and (\ref{eq:SCOT_Q}).

\subsection{Ground state energy}

\begin{figure}[t]
  \centering
  \includegraphics[width=0.5\textwidth,angle=270]{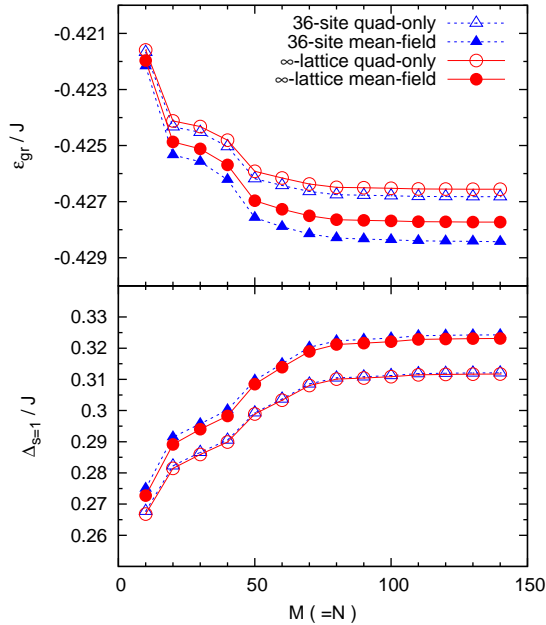}
  \caption{(Color online)
  The ground state energy per spin, $\epsilon_{gr}/J$, (top panel) and the lowest spin triplet excitation energy gap, $\Delta_{S=1}/J$, (bottom panel) as a function of spin triplet states included for the spin cluster operator representation.
  In this figure, the number of the included spin triplet states for the David star, $M$, and for each of the extended hexagons, $N$, are set to be equal and increased from 10 to 140.
  \label{fig:energy_spingap_map}}
\end{figure}

The ground state energy per spin is shown as a function of the spin-triplet state number included for the spin cluster operator representation in Fig.~\ref{fig:energy_spingap_map}, which also contains a similar plot for the spin triplet excitation gap.
Figure~\ref{fig:energy_spingap_map} shows computational results for four different situations:
(i) quadratic only approximation ignoring all quartic terms in the 36-site unit cell, denoted as ``36-site quad-only'' in the figure, (ii) self-consistent mean field decoupling of the quartic terms in the 36-site unit cell, denoted as ``36-site mean-field'',(iii) quadratic only approximation in the infinite lattice, denoted as ``$\infty$-lattice quad-only'', and (iv) self-consistent mean field decoupling in the infinite lattice, denoted as ``$\infty$-lattice mean-field''.  
As mentioned previously, the results begin to show a convergent behavior as the number of included spin triplet states in each spin cluster goes beyond about 80.

Table~\ref{tab:results} provides the fully self-consistent mean-field results at $M=N=140$ for the 36-site unit cell, denote as SCOT I$_{36}$, and the infinite lattice, denoted as SCOT I$_\infty$.
Note that, while the 36-site cluster has a little bit lower ground state energy  and a slightly higher spin triplet gap than the infinite lattice, the difference is very small compared to their absolute values. 
In the spin cluster operator theory, the ground state energy is estimated to be $-0.428 J$ per spin for the 36-site unit cell while it is given as $-0.438 J$ from exact diagonalization. 
Considering the crude nature of the mean-field approximation, we believe that such difference in the ground state energy is not too bad. 
Note that the energy difference is only $2.3 \%$ of the ground state energy itself.

Also, it is shown in Table~\ref{tab:results} that the SCOT I$_\infty$ ground state energy is lower than that of the BOT I$_\infty$ by $0.006 J$, which means that the spin cluster operator theory does a better job in utilizing the inter-dimer interaction to minimize the energy. 
In fact, the BOT I$_\infty$ simply reduces to the bond operator theory for a single spin cluster, E3, described in Table~\ref{tab:cluster_series_ED}.
This is because, in the bond operator mean field theory, any topologically orthogonal inter-dimer link behaves as if the link itself is absent.

\begin{table}[b]
\caption{
The ground state energy per spin, $\epsilon_{gr}/J$, and the lowest spin triplet excitation energy gap, $\Delta_{S=1}/J$, obtained from various computations.
The SCOT I and II denote the results obtained from the spin cluster operator theory for the VBS order with staggered and uniform hexagonal resonance, respectively.
Results from the 36-site unit cell and the infinite lattice are distinguished by the subscripts under I and II.
Similarly, the BOT I and II address the bond operator theory for the corresponding VBS orders.
Note that, while the BOT II$_\infty$ is essentially identical to the previous work done by some of the authors~\cite{Kagome_BOT}, there is a slight technical difference in the quadratic-decoupling procedure resulting in a minor quantitative modification.  
\label{tab:results}}
\begin{ruledtabular}
\begin{tabular}{lcc}
  &
  $
  \epsilon_{gr}/J
  $
  &
  $
  \Delta_{S=1}/J
  $
  \\
  \hline
  ED (36 sites)\cite{36-site_ED_Leung_Elser,36_ED_first_excited}&-0.438 &0.164
  \\
  SCOT I$_{36}$ &-0.4284 &0.3243
  \\
  SCOT II$_{36}$ &-0.4231 &0.2431
  \\
  \hline
  DMRG \cite{JiangWengSheng} &- &0.05
  \\
  BOT I$_\infty$ &-0.4221 &0.8327
  \\
  BOT II$_\infty$ &-0.4223 &0.7761
  \\
  SCOT I$_\infty$ &-0.4277 &0.3231
  \\
  SCOT II$_\infty$ &-0.4223 &0.2431
\end{tabular}
\end{ruledtabular}
\end{table}

\subsection{Spin triplet excitation spectrum}

\begin{figure*}
  \centering
  \includegraphics[width=1.0\linewidth]{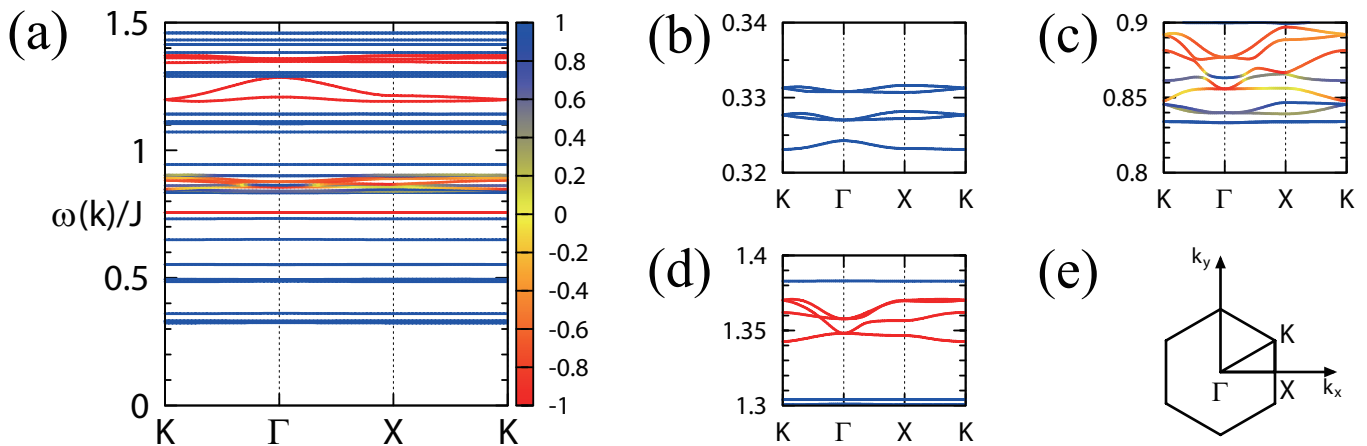}
  \caption{(Color online) Spin triplet excitation spectrum obtained from the spin cluster operator theory for the VBS order with staggered hexagonal resonance. 
  Figure (a) shows the triplet excitation spectrum with energy up to $1.5 J$. 
  Each spin triplet dispersion is color-coded depending on which region between the David star and the surrounding extended hexagons is more responsible for the origin of the corresponding spin triplet excitation.  
  The blue color indicates that the spin triplet excitation comes from the David star while the red color denotes otherwise.
 The yellow color indicates that there is an equal mixing between contributions from the both regions.
 Figure (b)-(d) show an enlarged view of the spectrum for the energy window of $\omega/J = 0.32 - 0.34$, $0.8 - 0.9$, and $1.3 - 1.4$, respectively. 
 Figure (e) shows the plotted path of the spin triplet excitation dispersion in the first Brillouin zone.
  \label{fig:band_I}}
\end{figure*}

Figure~\ref{fig:band_I} shows the spin triplet excitation spectrum of the spin cluster operator theory for the VBS order with staggered hexagonal resonance. 
As mentioned previously, the results presented in Fig.~\ref{fig:band_I} are those obtained at the full convergence occurring when the total number of the included spin triplet states for the spin cluster operator representation reaches $M+2N=420$ with $M=N=140$.

Before discussing details of the spin triplet excitation spectrum, it is convenient to introduce the following quantity characterizing the regional origin of each spin triplet excitation:
\begin{align}
 C_{l}({\bf k})
 =
 \frac{ \sum_{n=1}^{M} \left| [\mathrm{\bf T}^{-1}({\bf k})]_{nl} \right|^2 - \sum_{n=M+1}^{M+2N} \left| [\mathrm{\bf T}^{-1}({\bf k})]_{nl} \right|^2}{\sum_{n=1}^{M+2N} \left| [\mathrm{\bf T}^{-1}({\bf k})]_{nl} \right|^2}, 
\end{align}
where $l=1,\cdots,M+2N$ and $\mathrm{\bf T}^{-1}({\bf k})$ is the inverse Bogoliubov transformation matrix.
Noting that the spin triplet excitations from the David star are enumerated with indices between $1$ and $M$ and those from the extended hexagons are with indices between $M+1$ and $M+2N$,
the above quantity measures which region, the David star or the extended hexagons, is more responsible for a given spin triplet excitation. 
In the above definition, anomalous components of the eigenvectors, $[\mathrm{\bf T}^{-1}({\bf k})]_{nl} ~ (n=(M+2N)+1,\cdots,2(M+2N))$, are not included due to their little contribution.
Each dispersion curve in Fig.~\ref{fig:band_I} is color-coded based on the value of $C_l({\bf k})$ with the blue indicating that the corresponding spin triplet excitation comes from the David star and with the red otherwise.
The yellow color denotes that two contributions are almost equal.

As one can see from Fig.~\ref{fig:band_I} (a), most spin triplet excitation dispersions are nearly flat with few exceptions of the dispersive modes in high energy.
While there is in general non-zero mixing between the spin triplet modes originating from the David star and the surrounding extended hexagons,
low energy spin excitations mostly come from the David star which is more or less disconnected from the outside as indicated by the blue color.
In high energy, only a few dispersive modes, denoted in red, have dominant contributions from the extended hexagons. 
It is interesting to note that the lowest spin triplet dispersion in red are twofold degenerate and occurs around $0.75 J$ with a flat dispersion.
A similar behavior is observed in the case of the VBS order with uniform hexagonal resonance as shown in next section.

In addition to the blue and red spin triplet excitations exclusively originating from either region of the David star or the extended hexagons, 
spin triplet excitations in yellow have the equal contribution from both regions, appearing between $0.8$ and $0.9 J$.
Figure~\ref{fig:band_I} (c) shows an enlarged view of the spin triplet excitation spectrum in such region.
Roughly speaking, the spectrum becomes a superposition of the dispersive red and the relatively flat blue curve only when the two colored bands intersect each other. 
Away from the intersecting points, the spin triplet excitations possess mostly the single color characteristics.

The lowest triplet band is contained in the Fig.~\ref{fig:band_I} (b) together with the next four bands.
The lowest spin triplet excitation occurs at the K point with energy of $0.323 J$. 
It is interesting to compare the SCOT result for the lowest spin excitation with that of the usual bond operator theory.
The spin cluster operator theory predicts a much smaller spin triplet excitation gap than the bond operator theory. 
See Table~\ref{tab:results} for details.
Despite the better agreement, there still remains a factor of $3-6$ difference in comparison with the numerical results obtained from exact diagonalization as well as the density matrix renormalization group method.
While this difference could be induced by the crude approximation in the mean field theory itself,
we believe that it reveals a more fundamental limitation of the current spin cluster operator theory.
As mentioned in the above, the lowest spin excitation mostly originates from the David star.
Expecting that the true low-energy spin excitation has more or less the same contributions from both the David star and the extended hexagons, we believe that it is necessary to devise a theoretical framework where the spin triplet excitations from the both contributions are treated on a more equal footing.    
We discuss such theoretical framework in Sec.~\ref{sec:DISCUSSION}.

\section{RESULTS: UNIFORM HEXAGONAL RESONANCE \label{sec:RESULTS_II}}

\subsection{Ground state energy}

The ground state energy per spin for the VBS order with uniform hexagonal resonance is obtained as $-0.422 J$ in the fully self-consistent mean-field approximation, reduced from the quadratic-only approximation value, $-0.414 J$.
The SCOT result is in fact precisely identical to that of the bond operator theory as shown in Table~\ref{tab:results}.
The reason is that, in the VBS oder with uniform hexagonal resonance, the core cluster is completely decoupled from the surrounding dimers. 
It is mentioned in Sec.~\ref{subsec:topologically_orthogonal_vs_empty_triangle} that such decoupled core cluster does not have any quantum fluctuation about the pure dimer-covering VBS order and therefore the lowest energy per spin for the core cluster is simply given by the half energy of a spin singlet, $-0.375 J$.
Remaining contributions to the ground state energy are generated through the inter-dimer interaction between the surrounding dimers. 
These contributions, however, are exactly the same between the spin cluster and the bond operator theory.

\subsection{Spin triplet excitation spectrum}

\begin{figure}[t]
  \centering
  \includegraphics[width=0.65\linewidth,angle=270]{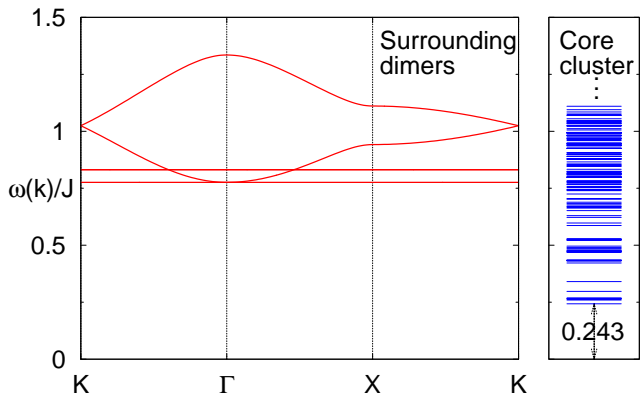}
  \caption{(Color online)
  Spin triplet excitation spectrum obtained from the spin cluster operator theory for the VBS order with uniform hexagonal resonance.   
  The spectrum consists of two separate sets of the spin triplet excitation with one coming from the core cluster and the other from the surrounding dimers. 
  The second lowest flat bands from the surrounding dimers are twofold degenerate.
  It is interesting to note that similar flat bands appear in the VBS order with staggered hexagon resonance with roughly the same energy.
  \label{fig:band_II}}
\end{figure}

The spin triplet excitation spectrum for the VBS order with uniform hexagonal resonance consists of two separate sets of excitations with one coming from the core cluster and the other from the surrounding dimers.
See Fig.~\ref{fig:band_II} for the spectrum of each set.  
Spin triplet excitations coming from the core cluster are entirely flat in dispersion due to the fact that the core cluster is completely decoupled from the surrounding dimers.
It is shown in Fig.~\ref{fig:band_II} that the lowest spin triplet excitation has an excitation energy gap of $0.243 J$.

The surrounding dimers provide nine spin triplet excitations with five of them being flat in dispersion.
Belonging to the five flat modes, the lowest spin triplet excitation has an energy of $0.776 J$ and is degenerate at the $\Gamma$ point with a dispersive mode.
Similar to the ground state energy, these results are identical to those of the bond operator theory since the connection between the core cluster and the surrounding dimers is absent.
The difference is that, in the current spin cluster operator theory, the actual lowest spin triplet excitation occurs from the core cluster instead of the surrounding dimers. 
For a similar reason mentioned previously in the case of the VBS order with staggered hexagonal resonance, we think that this separation between the core cluster and the surrounding dimers 
might be artificial.
To overcome this problem, in the next section, we propose a new theoretical framework for the spin cluster operator theory, where the both regions are treated in a more equal footing.

\section{DISCUSSION\label{sec:DISCUSSION}}

\begin{figure*}
  \centering
  \includegraphics[width=0.8\textwidth]{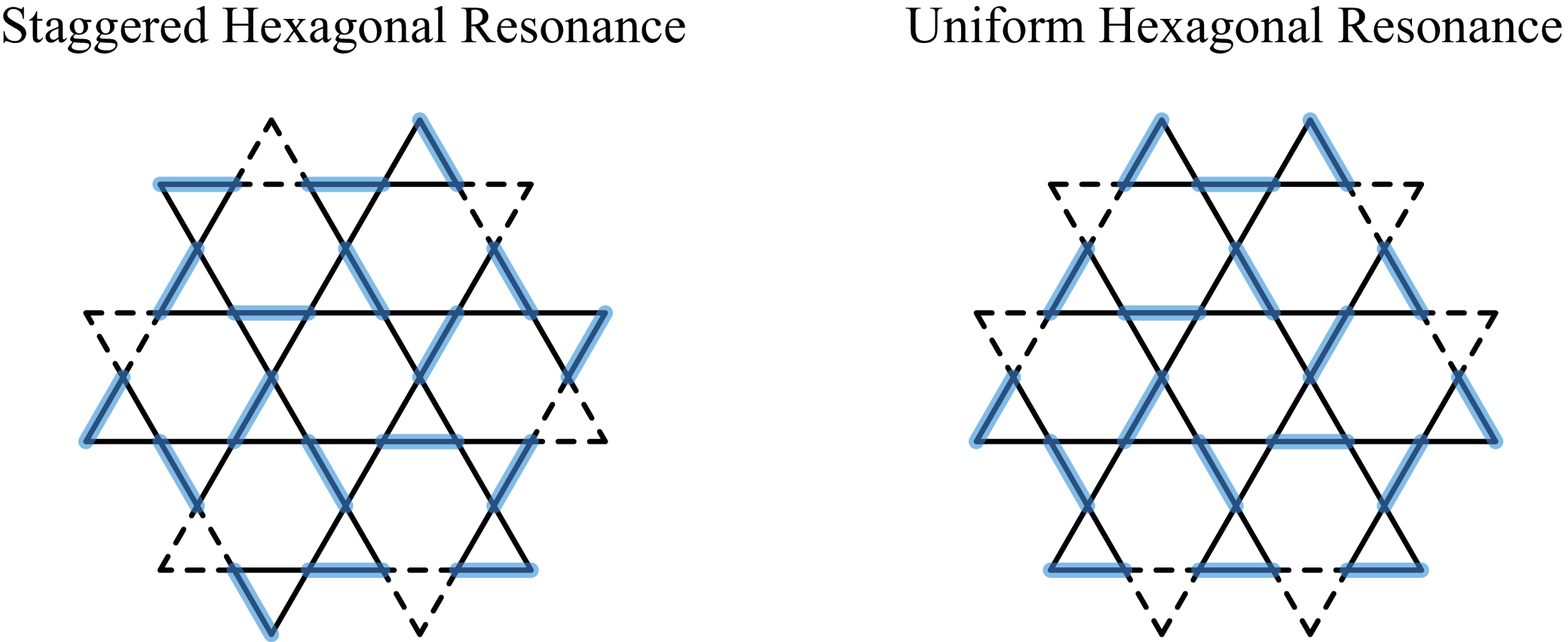}
  \caption{(Color online)
  Two dimer covering patterns of the super cluster for the VBS order with staggered (left panel) and uniform (right panel) hexagonal resonance. 
  Super cluster (black solid lines) is connected with its surroundings through the empty triangle inter-dimer interaction (dashed lines) only.
  In the figure, thick (blue) lines denote the spin singlet pairs in the VBS order.
  \label{fig:super_cluster}}
\end{figure*}

The spin cluster operator theory predicts a spin triplet excitation gap significantly reduced from that of the bond operator theory. 
This reduction results from the fact that effects of the topologically orthogonal inter-dimer interaction, which were entirely ignored in the bond operator theory, are better captured by the spin cluster operator theory.
Despite the improvement, however, the current spin cluster operator theory is not accurate enough to correctly describe the full effects of the topologically orthogonal inter-dimer interaction between dimers across different spin clusters. 
One way to see the inaccuracy is to realize that the spin triplet excitation gap is actually increased after the inter-dimer interactions between the David star and the extended hexagons are included in the VBS order with staggered hexagonal resonance.
This is in contradiction with the observation that spin clusters with more topologically orthogonal inter-dimer interactions generate a smaller spin triplet excitation gap.

An apparent source of the problem is that spin clusters considered in our current formalism do not fully take into account the effects of all topologically orthogonal inter-dimer interactions.
A natural conclusion based on this reasoning is that the best way to overcome this problem is to construct a ``super cluster'' that contains all topologically orthogonal inter-dimer interactions inside the unit cell.   
See Fig.~\ref{fig:super_cluster} for the illustration of possible super clusters.
Note that, similar to the David star in the VBS order with staggered hexagonal resonance and the core cluster in the uniform hexagonal resonance, two super clusters corresponding to each hexagonal resonance have twofold degeneracies with respect to the dimer covering orientation in the central pinwheel structure.

It is mentioned in Sec.~\ref{subsec:topologically_orthogonal_vs_empty_triangle} that there is absolutely no quantum fluctuation in the VBS state when all inter-dimer interactions are topologically orthogonal.
That is, the VBS state is a precise energy eigenstate of a spin cluster entirely composed of the topologically orthogonal inter-dimer interactions, in which case the energy of the VBS state is exactly given by $-0.375 J$ per spin.
Actually, it is shown in Table~\ref{tab:cluster_series_ED} that the VBS state is not only a precise energy eigenstate, but also the exact ground state for all finite-size spin clusters studied in this paper, (T1)-(T6).
Another important result obtained in Table~\ref{tab:cluster_series_ED} is that the lowest spin triplet excitation gap decreases as the number of topologically orthogonal inter-dimer interactions increases. 
Presuming that the same trend remains valid up to the super clusters,
we expect that the spin triplet excitation energy gap for the super cluster gets smaller than that of (T6), $0.243 J$, and becomes similar to that of the 36-site ED result with periodic boundary condition, $0.162 J$.

Covering entire topologically orthogonal inter-dimer interactions in the 36-site unit cell, the super cluster is connected with its surroundings through empty triangle inter-dimer interactions (dashed lines in Fig.~\ref{fig:super_cluster}) only. 
As mentioned in Sec.~\ref{subsec:topologically_orthogonal_vs_empty_triangle}, the main role of the empty triangle inter-dimer interaction is to minimize the ground state energy.
Considering that the energy-lowering effect of the empty triangle inter-dimer interaction has been rather well captured by the bond operator mean-field theory,
it is reasonable to assume that the same is true for the spin cluster operator mean-field theory.
For this reason, for future work, we propose to analyze the Kagome lattice antiferromagnetic Heisenberg model via the spin cluster operator theory with super clusters.

\section{SUMMARY \label{sec:SUMMARY}}

In this paper, we have considered the valence bond solid state with a 36-site unit cell as a possible ground state for the antiferromagnetic Heisenberg model on the Kagome lattice.
Our study is motivated by the realization that there is absolutely no quantum fluctuation about the spin-singlet product state when all inter-dimer interactions are topologically orthogonal. 
Assuming that the ground state tries to minimize the quantum fluctuation by forming as many topologically orthogonal inter-dimer interactions as possible, we arrive at the valence bond solid state with a 36-site unit cell.
The valence bond solid state has been analyzed previously by the bond operator theory where quantum modes emerging from a bond connecting two nearest neighboring spins are regarded as a natural building block for the dynamics of the system. 
While the bond operator theory is able to correctly reproduce the valence bond solid order in the ground state, it is shown that the spin triplet excitation spectrum is grossly overestimated in energy.

To diagnose the problem, we have carefully investigated effects of the two inter-dimer interaction structures, each named as the topologically orthogonal and the empty triangle structure. 
From exact diagonalization of various finite-sized spin clusters, it is learned that one has to treat the topologically orthogonal inter-dimer interaction in a more accurate manner than in the bond operator mean-field theory.
As a first step toward the accurate description for the topologically orthogonal inter-dimer interaction, the spin cluster operator theory is developed, in which natural quantum modes of the system are chosen from the spin clusters containing as many topologically orthogonal inter-dimer interactions as possible while preserving the symmetry of the VBS order ground state. 
In this paper, the spin cluster operator theory is used to analyze two different VBS states with one having the staggered hexagonal resonance and the other the uniform counterpart. 

As a result, it is shown that the lowest spin triplet excitation gap is much reduced from that of the bond operator theory, closely approaching the numerical estimate obtained via exact diagonalization as well as the density matrix renormalization group method.
Despite the better agreement, the fact that the lowest spin triplet excitation mostly comes from the inner region of the unit cell involving the pinwheel structure may be an artifact of the current spin cluster operator theory that uses a subset of the unit cell as a spin cluster.
For future work, we propose to use super spin clusters covering the entire unit cell for the spin cluster operator representation

\begin{acknowledgments}
This research was supported in part by the Korea Science and Engineering Foundation grant funded by the Korea government (MEST) under Quantum Metamaterials Research Center (K.P.).
K.H. is grateful to Heungsik Kim for his valuable help in setting  computational nodes.
Also, K.P. would like to thank Asia Pacific Center for Theoretical Physics (APCTP) for its hospitality. 
\end{acknowledgments}

\appendix
\section{Mean field decoupling of the quartic term
\label{appendix:Mean-field decoupling}}

In this section, we describe how to decouple the quartic interaction term between spin triplet bond particles into a mean-field quadratic Hamiltonian. 
Restoring the summation symbol, the usual quartic term of the spin triplet bond operator can be written
as follows:
\begin{align}
 &- \sum_{\alpha} \sum_{\beta \gamma} \sum_{\mu \nu} \epsilon_{\alpha\beta\gamma}\epsilon_{\alpha\mu\nu} t_{1\beta}^{\dagger}t_{1\gamma} t_{2\mu}^{\dagger}t_{2\nu}
 \nonumber\\
 &=
 - \sum_{\beta\ne\gamma} \left( t_{1\beta}^{\dagger} t_{2\beta}^{\dagger} t_{1\gamma} t_{2\gamma} - t_{1\beta}^{\dagger} t_{2\beta} t_{1\gamma} t_{2\gamma}^{\dagger} \right)
 \nonumber\\
 &= -\sum_{\beta\ne\gamma} \left( \langle t_{1\beta}^{\dagger} t_{2\beta}^{\dagger} \rangle t_{1\gamma} t_{2\gamma} + t_{1\beta}^{\dagger} t_{2\beta}^{\dagger} \langle t_{1\gamma} t_{2\gamma} \rangle \right.
 \nonumber\\
 &~~~~~~- \left. \langle t_{1\beta}^{\dagger} t_{2\beta}^{\dagger} \rangle \langle t_{1\gamma} t_{2\gamma} \rangle \right)
 \nonumber\\
 &+ \sum_{\beta\ne\gamma} \left( \langle t_{1\beta}^{\dagger} t_{2\beta} \rangle t_{1\gamma} t_{2\gamma}^{\dagger} + t_{1\beta}^{\dagger} t_{2\beta} \langle t_{1\gamma} t_{2\gamma}^{\dagger} \rangle \right.
 \nonumber\\
 &~~~~~~ - \left. \langle t_{1\beta}^{\dagger} t_{2\beta} \rangle \langle t_{1\gamma} t_{2\gamma}^{\dagger} \rangle \right)
 \nonumber\\
 &=~~\frac{2}{3}(Q^{*}Q-P^{*}P)
 \nonumber\\
 &+\frac{2}{3}\sum_{\alpha} \left( P t_{1\alpha}^{\dagger} t_{2\alpha} - Q t_{1\alpha}^{\dagger} t_{2\alpha}^{\dagger} +\textrm{H. c.} \right) ,
 \label{eq:mean-field decoupling}
\end{align}
where $P= \langle t_{1\alpha} t^\dagger_{2\alpha} \rangle$ and $Q=\langle t_{1\alpha} t_{2\alpha}\rangle$.
Note that, in the first step of the above equation, we have used the identity that $\sum_{\alpha}\epsilon_{\alpha\beta\gamma}\epsilon_{\alpha\mu\nu}=\delta_{\beta\mu}\delta_{\gamma\nu}-\delta_{\beta\nu}\delta_{\gamma\mu}$.
\\
\\ 

\section{Bond operator theory in the empty triangle structure
\label{appendix:BOT_in_ET}}

The Heisenberg-model Hamiltonian for the empty triangle unit structure, E1, in Fig.~\ref{fig:inter_dimer_int} is given by
\begin{eqnarray}
 H_{\textrm{E1},J} &=&  J \left( {\bf S}_a \cdot {\bf S}_b + {\bf S}_c \cdot {\bf S}_d + {\bf S}_e \cdot {\bf S}_f \right)
 \nonumber\\ 
 &+& \lambda J \left( {\bf S}_b \cdot {\bf S}_d + {\bf S}_d \cdot {\bf S}_f + {\bf S}_f \cdot {\bf S}_b \right) .
 \label{eq:H_E1_J}
\end{eqnarray}
Performing essentially the same procedure in the analysis of the topologically orthogonal structure, 
the bond operator mean field Hamiltonian can be obtained as follows:  
\begin{align}
 H^\textrm{MF}_\textrm{E1} &= \epsilon_{\textrm{E1},0} + \left( \frac{J}{4} - \mu \right) \sum_{i=1}^{3} t_{i\alpha}^{\dagger} t_{i\alpha} 
 \nonumber\\
 &+ \lambda J \sum_{i=1}^{3} \left( A t_{i\alpha}^{\dagger} t_{i+1\alpha} + B t_{i\alpha}^{\dagger} t_{i+1\alpha}^{\dagger}  + \textrm{H. c.}\right),
 \label{eq:H_MF_E1}
\end{align}
where
\begin{align}
\epsilon_{\textrm{E1},0} = 3 \left[ -\frac{3}{4} J {\bar s}^2 + \mu (1-{\bar s}^2) 
- \frac{\lambda J}{4} \cdot \frac{2}{3}(P^{*}P-Q^{*}Q) \right], 
\end{align}
\begin{align} 
A &= \frac{1}{4} \left( {\bar s}^2 + \frac{2}{3}P \right),
\\
B &= \frac{1}{4} \left( {\bar s}^2 - \frac{2}{3}Q \right).
\end{align}
Note that, while there are three different bonds in E1, we set the spin-singlet condensation density to be identical across all the bonds with
the same being true for the corresponding chemical potential. 
That is, $s_1=s_2=s_3={\bar s}$ and $\mu_1=\mu_2=\mu_3=\mu$.
In the above, the mean field parameters, $P$ and $Q$, are defined as follows:
\begin{eqnarray}
 P &=& \left< t_{i\alpha} t_{i+1,\alpha}^{\dagger} \right>,
 \\
 Q &=& \left< t_{i\alpha} t_{i+1,\alpha} \right>,
 \label{eq:PQ_E1}
\end{eqnarray}
where $t_{i+3,\alpha}$ is equal to $t_{i\alpha}$.

The mean field Hamiltonian, $H^\textrm{MF}_\textrm{E1}$, is block-diagonalized in a proper basis set that takes into account the threefold rotational symmetry of the system.
The new basis set of the triplet operator is defined as follows: 
\begin{eqnarray}
 {\tilde t}_{m\alpha} = \frac{1}{\sqrt{3}} \sum_{i=1}^3 (z_m)^{i-1} t_{i\alpha},
 \label{eq:basis_change}
\end{eqnarray} 
where $z_m=e^{i \frac{2\pi}{3} m}$ and $m=-1,0,1$.
In terms of this basis set, the mean field Hamiltonian is given by
\begin{align}
 H^\textrm{MF}_\textrm{E1} &=  \epsilon_{\textrm{E1},0} - \frac{9}{2} \left( \frac{J}{4} -\mu \right)
 \nonumber\\
 &+\frac{1}{2} 
 \sum_{m=-1}^{1}
 \left(
 \begin{array}{cc}
  \tilde{t}_{m\alpha}^{\dagger}
  &
  \tilde{t}_{-m\alpha}
 \end{array}
 \right)
 \left(
 \begin{array}{cc}
 A_m & B_m
 \\
 B_m^* & A_{-m} 
 \end{array}
 \right)
 \left(
 \begin{array}{c}
  \tilde{t}_{m\alpha}
  \\
  \tilde{t}_{-m\alpha}^{\dagger}
 \end{array}
 \right),
 \nonumber\\
 \label{eq:ET_H_MF}
\end{align}
where
\begin{align}
 A_m &= \frac{J}{4} - \mu + \lambda J (z_{-m} A + z_m A^*),
 \\
 B_m &= \lambda J (z_m + z_{-m}) B.
\end{align}

A complete diagonalization of $H^\textrm{MF}_\textrm{E1}$ is 
performed via the Bogoliubov transformation: 
\begin{align}
H^\textrm{MF}_\textrm{E1} &=  \epsilon_{gr,\textrm{E1}} +\sum_{m=-1}^{1} \omega_m \gamma_m^{\dagger} \gamma_m \;,
\end{align}
where $\omega_m$ is the positive eigenvalues of the following eigenvalue problem:
\begin{align}
 &
 \left(
 \begin{array}{cc}
 A_m & B_m
 \\
 -B_m^* & -A_{-m} 
 \end{array}
 \right)
 \left(
 \begin{array}{cc}
 u_m^{(1)} & u_m^{(2)} 
 \\
 v_m^{(1)} & v_m^{(2)} 
 \end{array}
 \right)
 \nonumber\\
 &=
 \left(
 \begin{array}{cc}
 u_m^{(1)} & u_m^{(2)} 
 \\
 v_m^{(1)} & v_m^{(2)} 
 \end{array}
 \right)
 \left(
 \begin{array}{cc}
 \omega_m & 0 
 \\
 0 & -\omega_m 
 \end{array}
 \right) \;,
\end{align}
and
\begin{eqnarray}
\epsilon_{gr,\textrm{E1}} = \epsilon_\textrm{E1,0} - \frac{9}{2}\left(\frac{J}{4}-\mu\right) + \frac{3}{2} \sum_{m=-1}^{1} \omega_m \,.
\end{eqnarray}
Note that the Bogoliubov quasi-particle operator is transformed to the original triplet bond operator through the transformation matrix composed of the eigenvectors in the above:
\begin{eqnarray}
 \left(
 \begin{array}{c}
  \tilde{t}_{m\alpha}
  \\
  \tilde{t}_{-m\alpha}^{\dagger}
 \end{array}
 \right)
 =
 \left(
 \begin{array}{cc}
 u_m^{(1)} & u_m^{(2)} 
 \\
 v_m^{(1)} & v_m^{(2)} 
 \end{array}
 \right)
 \left(
 \begin{array}{c}
  \gamma_{m\alpha}
  \\
  \gamma_{-m\alpha}^{\dagger}
 \end{array}
 \right) ,
 \label{eq:ET_trans}
\end{eqnarray}
where $u^{(1,2)}_m$ and $v^{(1,2)}_m$ are normalized in such a way that
\begin{eqnarray}
 && |u_m^{(1)}|^2 - |v_m^{(1)}|^2 = 1,
 \\
 && |u_m^{(2)}|^2 - |v_m^{(2)}|^2 = -1,
 \\
 && u_m^{(1)*} u_m^{(2)} - v_m^{(1)*} v_m^{(2)} = 0 \;.
\end{eqnarray}

The spin-spin correlation value can be computed for the $(a,b)$ and $(b,d)$ link in the ground state as follows:  
\begin{eqnarray}
 \left< {\bf S}_a \cdot {\bf S}_b \right>
 &=&  \leftidx{_\textrm{E1}}{\Big< \textrm{gr} \Big|} \left[ -\frac{3}{4} s_1^{\dagger} s_1 + \frac{1}{4} \sum_\alpha t_{1\alpha}^{\dagger} t_{1\alpha} \right] \Big| \textrm{gr} \Big>_\textrm{E1}
 \nonumber\\
 &=& -\frac{3}{4} {\bar s}^2 + \frac{1}{4} (1-{\bar s}^2)
 = \frac{1}{4} - {\bar s}^2
\end{eqnarray}
and
\begin{eqnarray}
 \left< {\bf S}_b \cdot {\bf S}_d \right>
 &=& \frac{1}{3J} \left< H_{\textrm{E1},J} \right> - \left< {\bf S}_a \cdot {\bf S}_b \right>
 \label{eq:spin_corr}
 \\
 &=& \frac{1}{3J} \epsilon_{gr,\textrm{E1}} - \left( \frac{1}{4} - {\bar s}^2 \right).
\end{eqnarray}
The spin-spin correlation value for the other links can be determined through the space rotation symmetry.

The spin-spin correlation value for the excited state can be computed in a similar manner:
\begin{align}
 \left< {\bf S}_a \cdot {\bf S}_b \right>
 &= 
 \leftidx{_\textrm{E1}} {\Big< \textrm{gr} \Big|}  
 \gamma_{m\alpha} \left[ -\frac{3}{4} s_1^{\dagger} s_1 + \frac{1}{4} \sum_{\alpha'} t_{1\alpha'}^{\dagger} t_{1\alpha'} \right] \gamma_{m\alpha}^{\dagger} 
 \Big| \textrm{gr} \Big>_\textrm{E1}
  \nonumber\\
 &= -\frac{3}{4} {\bar s}^2 + \frac{1}{4} \sum_{\alpha'} \leftidx{_\textrm{E1}}{\left< \textrm{gr} \right|} \gamma_{m\alpha} t_{1\alpha'}^{\dagger} t_{1\alpha'} \gamma_{m\alpha}^{\dagger} \left| \textrm{gr} \right>_\textrm{E1} \;,
\end{align}
where the last term is evaluated by using Eq.~(\ref{eq:basis_change}) and (\ref{eq:ET_trans}) as well as the fact that $\gamma_{m\alpha} \left| \textrm{gr} \right>_\textrm{E1} = 0$: 
\begin{eqnarray}
 &&\leftidx{_\textrm{E1}}{\left< \textrm{gr} \right|} \gamma_{m\alpha} t_{1\alpha'}^{\dagger} t_{1\alpha'} \gamma_{m\alpha}^{\dagger} \left| \textrm{gr} \right>_\textrm{E1}
 \nonumber\\
 &=&
 \frac{1}{3} \left( \left| u_m^{(1)} \right|^2 + \left| u_{-m}^{(2)} \right|^2 + 3 \sum_{l=-1}^{1} \left| u_{l}^{(2)} \right|^2 \right).
\end{eqnarray}
Finally, $\left< {\bf S}_b \cdot {\bf S}_d \right>$ can be obtained in a similar manner described in Eq.~(\ref{eq:spin_corr}):
\begin{eqnarray}
 \left< {\bf S}_b \cdot {\bf S}_d \right>
  = \frac{1}{3J} \left( \epsilon_{gr,\textrm{E1}} +\omega_{m} \right)
  - \left< {\bf S}_a \cdot {\bf S}_b \right>.
\end{eqnarray}
\\
\\

\section{Brief sketch for the derivation of the spin cluster operator representation
\label{appendix:SCOT_representation}}

The spin cluster operator representation shown in Eq.~(\ref{eq:David_representation}) is a result of the simplification due to a spin rotation symmetry.
With the spin rotation operator defined as
\begin{eqnarray}
\mathrm{R}_{\hat{n}}(\phi) = e^{ -i \phi \hat{n} \cdot {\bf S} },
\end{eqnarray}
the spin triplet states satisfy the following relationships:
\begin{subequations}
\begin{eqnarray}
 \mathrm{R}_{\hat{\alpha}} \left( \frac{\pi}{2} \right) \left| t_\beta \right> &=& 
 \left\lbrace
 \begin{array}{cc}
  \epsilon_{\alpha \beta \gamma} \left| t_\gamma \right> & \left( \alpha \ne \beta \right)
  \\
  \left| t_\beta \right> & \left( \alpha = \beta \right)
 \end{array}
 \right., 
 \label{eq:pi_over_2_state}
 \\
 \mathrm{R}_{\hat{\alpha}} \left( \pi \right) \left| t_\beta \right> &=& 
 \left\lbrace
 \begin{array}{cc}
  - \left| t_\beta \right> & \left( \alpha \ne \beta \right)
  \\
  \left| t_\beta \right> & \left( \alpha = \beta \right)
 \end{array}
 \right.,
 \label{eq:pi_state}
\end{eqnarray}
\end{subequations}
where $\alpha,\beta,\gamma \in \left\lbrace x,y,z \right\rbrace.$ 
Meanwhile, the spin rotation operator annihilates the spin singlet state, $\left| s \right>$:
\begin{eqnarray}
 \mathrm{R}_{\hat{n}}(\phi) \left| s \right> = \left| s \right>.
 \label{eq:singlet_transf}
\end{eqnarray}
Also, the spin operator itself satisfies the following transformation rules:
\begin{subequations}
\begin{eqnarray}
 \mathrm{R}_{\hat{\alpha}} \left( \frac{\pi}{2} \right) \mathrm{S}_{n\beta} \mathrm{R}_{\hat{\alpha}}^{\dagger} \left( \frac{\pi}{2} \right) &=& 
 \left\lbrace
 \begin{array}{cc}
  \epsilon_{\alpha \beta \gamma} \mathrm{S}_{n\gamma} & \left( \alpha \ne \beta \right)
  \\
  \mathrm{S}_{n\beta} & \left( \alpha = \beta \right)
 \end{array}
 \right.,
 \label{eq:pi_over_2_op} 
 \\
 \mathrm{R}_{\hat{\alpha}} \left( \pi \right) \mathrm{S}_{n\beta} \mathrm{R}_{\hat{\alpha}}^{\dagger} \left( \pi \right) &=& 
 \left\lbrace
 \begin{array}{cc}
  - \mathrm{S}_{n\beta} & \left( \alpha \ne \beta \right)
  \\
  \mathrm{S}_{n\beta} & \left( \alpha = \beta \right)
 \end{array}
 \right. .
 \label{eq:pi_op}
\end{eqnarray}
\end{subequations}

Making use of the above results, one can derive a series of general properties for the matrix element of the spin operator:
\begin{eqnarray}
 \label{eq:zero1}
 && \left< s \right| S_{n\alpha} \left| s \right> = 0,	\\
 \label{eq:zero2}
 && \left< s \right| S_{n\alpha} \left| t_{\beta} \right> = 0 \ \ (\alpha \ne \beta),	\\
 \label{eq:nonzero1}
 && \left< s \right| S_{nx} \left| t_{x} \right>
 = \left< s \right| S_{ny} \left| t_{y} \right>
 = \left< s \right| S_{nz} \left| t_{z} \right>.
\end{eqnarray}
For any two triplet states, $\left| t_{1\alpha} \right>$ and $\left| t_{2\beta} \right>$,
\begin{eqnarray}
 && \left< t_{1y} \right| S_{nx} \left| t_{2z} \right> = - \left< t_{1z} \right| S_{nx} \left| t_{2y} \right>  \nonumber\\
 &=& \left< t_{1z} \right| S_{ny} \left| t_{2x} \right> = - \left< t_{1x} \right| S_{ny} \left| t_{2z} \right>  \nonumber\\
 &=& \left< t_{1x} \right| S_{nz} \left| t_{2y} \right> = - \left< t_{1y} \right| S_{nz} \left| t_{2x} \right>. 
\label{eq:nonzero2}
\end{eqnarray}
Note that
\begin{equation}
 \left< t_{1\beta} \right| S_{n\alpha} \left| t_{2\gamma} \right> = 0
 \label{eq:zero3}
\end{equation}
if any two indices among $(\alpha,\beta,\gamma)$ are the same.
Finally, it is due to Eq.~(\ref{eq:nonzero2}) that the spin operator representation matrices in Eq.~(\ref{eq:e_mn}) and (\ref{eq:b_mn}) are anti-hermitian.
\\
\\

\section{Details of the mean field Hamiltonian in Section~\ref{sec:SCOT_I}
\label{appendix:details_of_the_mean_field_Hamiltonian}}

The mean field Hamiltonian (\ref{eq:H_{MF}}) has following form:
\begin{widetext}
\begin{eqnarray}
 && H_{MF}	\nonumber\\
 &=& \mathscr{N} \epsilon_o	+ \mathscr{N} \epsilon_{\mathscr{PQ}} \nonumber\\
 &+& \sum_{{\bf k}} \sum_{n=1}^{M} \left( \epsilon^{D}_{n} - \mu_{D} \right) t^{D\dagger}_{n\alpha}({\bf k}) t^{D}_{n\alpha}({\bf k})	\nonumber\\
 &+& \sum_{{\bf k}} \sum_{n=1}^{N} \left( \epsilon^{H}_{n} - \mu_{H} \right) t^{H1\dagger}_{n\alpha}({\bf k}) t^{H1}_{n\alpha}({\bf k})	\nonumber\\
 &+& \sum_{{\bf k}} \sum_{n=1}^{N} \left( \epsilon^{H}_{n} - \mu_{H} \right) t^{H2\dagger}_{n\alpha}({\bf k}) t^{H2}_{n\alpha}({\bf k})	\nonumber\\
 &+& \lambda J  \sum_{{\bf k}} \sum_{m=1}^{M} \sum_{n=1}^{N}
 (A_{mn}({\bf k})+\mathscr{A}_{mn}({\bf k}))
  t_{m\alpha}^{D\dagger}({\bf k}) t_{n\alpha}^{H1\dagger}(-{\bf k})
 + \textrm{H. c.}
 \nonumber\\
 &+& \lambda J  \sum_{{\bf k}} \sum_{m=1}^{M} \sum_{n=1}^{N}
 (B_{mn}({\bf k})+\mathscr{B}_{mn}({\bf k})) t_{m\alpha}^{D\dagger}({\bf k}) t_{n\alpha}^{H1}({\bf k})
 + \textrm{H. c.}
 \nonumber\\
 &+& \lambda J  \sum_{{\bf k}} \sum_{m=1}^{M} \sum_{n=1}^{N}
 (C_{mn}({\bf k})+\mathscr{C}_{mn}({\bf k})) t_{m\alpha}^{D\dagger}({\bf k}) t_{n\alpha}^{H2\dagger}(-{\bf k})
 + \textrm{H. c.}
 \nonumber\\
 &+& \lambda J  \sum_{{\bf k}} \sum_{m=1}^{M} \sum_{n=1}^{N}
 (D_{mn}({\bf k})+\mathscr{D}_{mn}({\bf k})) t_{m\alpha}^{D\dagger}({\bf k}) t_{n\alpha}^{H2}({\bf k})
 + \textrm{H. c.}
 \nonumber\\
 &+& \lambda J  \sum_{{\bf k}}  \sum_{m=1}^{N} \sum_{n=1}^{N}
 (R_{mn}({\bf k})+\mathscr{R}_{mn}({\bf k})) t_{m\alpha}^{H1\dagger}({\bf k}) t_{n\alpha}^{H2\dagger}(-{\bf k})
 + \textrm{H. c.}
 \nonumber\\
 &+& \lambda J  \sum_{{\bf k}}  \sum_{m=1}^{N} \sum_{n=1}^{N}
 (S_{mn}({\bf k})+\mathscr{S}_{mn}({\bf k})) t_{m\alpha}^{H1\dagger}({\bf k}) t_{n\alpha}^{H2}({\bf k})
 + \textrm{H. c.},
 \label{eq:Kagome_H_MF}
\end{eqnarray}
where $\epsilon^{D}_{n}$ ($\epsilon^{H}_{n}$) means the energy of triplet state $\left| t_{n\alpha} \right>$ in the David star (extended hexagon), and
$\epsilon_{o}$ is defined in (\ref{eq:SCOT_e_o}).
Remaining undefined quantities are defined below:

\begin{eqnarray}
 && \epsilon_{\mathscr{PQ}} \nonumber\\
 &=& \frac{2}{3} \lambda J \sum_{k,l=1}^{M} \sum_{p,q=1}^{N}
 e_{kl}^{(7)} ( b_{pq}^{(7)} + b_{pq}^{(8)} ) \nonumber\\
 && \left\{ \left[ \mathscr{P}_{kq}^{DH1\dagger}({\bf 0}) \cdot \mathscr{P}_{lp}^{DH1}({\bf 0})
 - \mathscr{Q}_{kp}^{DH1\dagger}({\bf 0}) \cdot \mathscr{Q}_{lq}^{DH1}({\bf 0}) \right] r_p^{2} r_q^{-2} \right. \nonumber\\
 && + \left[ \mathscr{P}_{kq}^{DH1\dagger}({\bf r}_{B}) \cdot \mathscr{P}_{lp}^{DH1}({\bf r}_{B})
 - \mathscr{Q}_{kp}^{DH1\dagger}({\bf r}_{B}) \cdot \mathscr{Q}_{lq}^{DH1}({\bf r}_{B}) \right] z_k^{2} z_l^{-2} \nonumber\\
 && + \left. \left[ \mathscr{P}_{kq}^{DH1\dagger}({\bf r}_{C}) \cdot \mathscr{P}_{lp}^{DH1}({\bf r}_{C})
 - \mathscr{Q}_{kp}^{DH1\dagger}({\bf r}_{C}) \cdot \mathscr{Q}_{lq}^{DH1}({\bf r}_{C}) \right] z_k^{4} z_l^{-4} r_p r_q^{-1} \right\} \nonumber\\
 &+& \frac{2}{3} \lambda J \sum_{k,l=1}^{M} \sum_{p,q=1}^{N}
 e_{kl}^{(7)} ( b_{pq}^{(7)} + b_{pq}^{(8)} ) \nonumber\\
 &\cdot& \left\{ \left[ \mathscr{P}_{kq}^{DH2\dagger}({\bf r}_{A}) \cdot \mathscr{P}_{lp}^{DH2}({\bf r}_{A})
 - \mathscr{Q}_{kp}^{DH2\dagger}({\bf r}_{A}) \cdot \mathscr{Q}_{lq}^{DH2}({\bf r}_{A}) \right] z_k z_l^{-1} \right.\nonumber\\
 && + \left[ \mathscr{P}_{kq}^{DH2\dagger}({\bf r}_{B}) \cdot \mathscr{P}_{lp}^{DH2}({\bf r}_{B})
 - \mathscr{Q}_{kp}^{DH2\dagger}({\bf r}_{B}) \cdot \mathscr{Q}_{lq}^{DH2}({\bf r}_{B}) \right] z_k^{3} z_l^{-3} r_p r_q^{-1} \nonumber\\
 && + \left. \left[ \mathscr{P}_{kq}^{DH2\dagger}({\bf 0}) \cdot \mathscr{P}_{lp}^{DH2}({\bf 0})
 - \mathscr{Q}_{kp}^{DH2\dagger}({\bf 0}) \cdot \mathscr{Q}_{lq}^{DH2}({\bf 0}) \right] z_k^{5} z_l^{-5} r_p^2 r_q^{-2} \right\} \nonumber\\
 &+& \frac{2}{3} \lambda J \sum_{k,l=1}^{M} \sum_{p,q=1}^{N}
 \left[ b_{kl}^{(8)} ( b_{pq}^{(2)} + b_{pq}^{(1)} r_p r_q^{-1} ) + b_{pq}^{(8)} ( b_{kl}^{(2)} + b_{kl}^{(1)} r_k r_l^{-1} )  \right] \nonumber\\
 &\cdot&  \left\{ \left[ \mathscr{P}_{kq}^{HH\dagger}(-{\bf r}_{C}) \cdot \mathscr{P}_{lp}^{HH}(-{\bf r}_{C})
 - \mathscr{Q}_{kp}^{HH\dagger}(-{\bf r}_{C}) \cdot \mathscr{Q}_{lq}^{HH}(-{\bf r}_{C}) \right] r_p^{2} r_q^{-2} \right. \nonumber\\
 &&  + \left[ \mathscr{P}_{kq}^{HH\dagger}({\bf r}_{A}) \cdot \mathscr{P}_{lp}^{HH}({\bf r}_{A})
 - \mathscr{Q}_{kp}^{HH\dagger}({\bf r}_{A}) \cdot \mathscr{Q}_{lq}^{HH}({\bf r}_{A}) \right] r_k r_l^{-1} \nonumber\\
 &&  \left. + \left[ \mathscr{P}_{kq}^{HH\dagger}({\bf 0}) \cdot \mathscr{P}_{lp}^{HH}({\bf 0})
 - \mathscr{Q}_{kp}^{HH\dagger}({\bf 0}) \cdot \mathscr{Q}_{lq}^{HH}({\bf 0}) \right] r_k^{2} r_l^{-2} r_p r_q^{-1} \right\},
\end{eqnarray}

\begin{subequations}
\begin{eqnarray}
 && A_{mn}({\bf k}) = {\bar s}_D {\bar s}_{H} f_{m}^{(7)} ( a_{n}^{(7)} + a_{n}^{(8)} ) ( r_n^{2} + z_m^{2} e^{-i{\bf k} \cdot {\bf r}_B} + z_m^{4} r_n e^{-i{\bf k} \cdot {\bf r}_C} ),	\\
 && B_{mn}({\bf k}) = {\bar s}_D {\bar s}_{H} f_{m}^{(7)} ( a_{n}^{(7)*} + a_{n}^{(8)*} ) ( r_n^{*2} + z_m^{2} e^{-i{\bf k} \cdot {\bf r}_B} + z_m^{4} r_n^{*} e^{-i{\bf k} \cdot {\bf r}_C} ),	\\
 && C_{mn}({\bf k}) = {\bar s}_D {\bar s}_{H} f_{m}^{(7)} ( a_{n}^{(7)} + a_{n}^{(8)} ) ( z_m e^{-i{\bf k} \cdot {\bf r}_A}+ z_m^{3} r_n e^{-i{\bf k} \cdot {\bf r}_B} + z_m^{5} r_n^{2} ),	\\
 && D_{mn}({\bf k}) = {\bar s}_D {\bar s}_{H} f_{m}^{(7)} ( a_{n}^{(7)*} + a_{n}^{(8)*} ) ( z_m e^{-i{\bf k} \cdot {\bf r}_A} + z_m^{3} r_n^{*} e^{-i{\bf k} \cdot {\bf r}_B} + z_m^{5} r_n^{*2} ),	\\
 && R_{mn}({\bf k}) = {\bar s}_{H} {\bar s}_{H} \left[ a_{m}^{(8)} ( a_{n}^{(2)} + a_{n}^{(1)} r_n) + a_{n}^{(8)} ( a_{m}^{(2)} + a_{m}^{(1)} r_m) \right]
 ( r_n^{2} e^{i{\bf k} \cdot {\bf r}_C} + r_m e^{-i{\bf k} \cdot {\bf r}_A} + r_m^{2} r_n), \\
 && S_{mn}({\bf k}) = {\bar s}_{H} {\bar s}_{H} \left[ a_{m}^{(8)} ( a_{n}^{(2)*} + a_{n}^{(1)*} r_n^{*} ) + a_{n}^{(8)*} ( a_{m}^{(2)} + a_{m}^{(1)} r_m ) \right]
 ( r_n^{*2} e^{i{\bf k} \cdot {\bf r}_C} + r_m e^{-i{\bf k} \cdot {\bf r}_A} + r_m^{2} r_n^{*} ),
\end{eqnarray}
\end{subequations}

%
\begin{subequations}
\begin{eqnarray}
  \mathscr{A}_{mn}({\bf k})
 &=& \frac{2}{3} \sum_{k=1}^{M} \sum_{l=1}^{N} e_{mk}^{(7)} ( b_{nl}^{(7)} + b_{nl}^{(8)} )	\nonumber\\
 && \cdot \left[ r_n^{2} r_l^{-2} \mathscr{Q}_{kl}^{DH1}({\bf 0})
 + z_m^{2} z_k^{-2} \mathscr{Q}_{kl}^{DH1}({\bf r}_B) e^{-i{\bf k}\cdot{\bf r}_{B}}
 + z_m^{4} z_k^{-4} r_n r_l^{-1} \mathscr{Q}_{kl}^{DH1}({\bf r}_C) e^{-i{\bf k}\cdot{\bf r}_{C}} \right],
\end{eqnarray}
\begin{eqnarray}
 \mathscr{B}_{mn}({\bf k})
 &=& - \frac{2}{3} \sum_{k=1}^{M} \sum_{l=1}^{N} e_{mk}^{(7)} ( b_{ln}^{(7)} + b_{ln}^{(8)} )	\nonumber\\
 && \cdot \left[ r_l^{2} r_n^{-2} \mathscr{P}_{kl}^{DH1}({\bf 0})
 + z_m^{2} z_k^{-2} \mathscr{P}_{kl}^{DH1}({\bf r}_B) e^{-i{\bf k}\cdot{\bf r}_{B}}
 + z_m^{4} z_k^{-4} r_l r_n^{-1} \mathscr{P}_{kl}^{DH1}({\bf r}_C) e^{-i{\bf k}\cdot{\bf r}_{C}} \right],
\end{eqnarray}
\begin{eqnarray}
 \mathscr{C}_{mn}({\bf k})
 &=& \frac{2}{3} \sum_{k=1}^{M} \sum_{l=1}^{N} e_{mk}^{(7)} ( b_{nl}^{(7)} + b_{nl}^{(8)} )	\nonumber\\
 && \cdot \left[ z_m z_k^{-1} \mathscr{Q}_{kl}^{DH2}({\bf r}_A)  e^{-i{\bf k}\cdot{\bf r}_A}
 + z_m^{3} z_k^{-3} r_n r_l^{-1} \mathscr{Q}_{kl}^{DH2}({\bf r}_B) e^{-i{\bf k}\cdot{\bf r}_B}
 + z_m^{5} z_k^{-5} r_n^2 r_l^{-2} \mathscr{Q}_{kl}^{DH2}({\bf 0}) \right],
\end{eqnarray}
\begin{eqnarray}
 \mathscr{D}_{mn}({\bf k})
 &=& - \frac{2}{3} \sum_{k=1}^{M} \sum_{l=1}^{N} e_{mk}^{(7)} ( b_{ln}^{(7)} + b_{ln}^{(8)} )	\nonumber\\
 && \cdot \left[ z_m z_k^{-1} \mathscr{P}_{kl}^{DH2}({\bf r}_A)  e^{-i{\bf k}\cdot{\bf r}_A}
 + z_m^{3} z_k^{-3} r_l r_n^{-1} \mathscr{P}_{kl}^{DH2}({\bf r}_B) e^{-i{\bf k}\cdot{\bf r}_B}
 + z_m^{5} z_k^{-5} r_l^2 r_n^{-2} \mathscr{P}_{kl}^{DH2}({\bf 0}) \right],
\end{eqnarray}
\begin{eqnarray}
 \mathscr{R}_{mn}({\bf k})
 &=& \frac{2}{3} \sum_{k=1}^{N} \sum_{l=1}^{N}
 \left[
 b_{mk}^{(8)} ( b_{nl}^{(2)} + b_{nl}^{(1)} r_n r_l^{-1} )
 + b_{nl}^{(8)} ( b_{mk}^{(2)} + b_{mk}^{(1)} r_m r_k^{-1} )
 \right] 	\nonumber\\
 && \cdot \left[ r_n^2 r_l^{-2} \mathscr{Q}_{kl}^{HH}(-{\bf r}_C)  e^{i{\bf k}\cdot{\bf r}_C}
 + r_m r_k^{-1} \mathscr{Q}_{kl}^{HH}({\bf r}_A) e^{-i{\bf k}\cdot{\bf r}_A}
 + r_m^{2} r_k^{-2} r_n r_l^{-1} \mathscr{Q}_{kl}^{HH}({\bf 0}) \right],
\end{eqnarray}
\begin{eqnarray}
 \mathscr{S}_{mn}({\bf k})
 &=& - \frac{2}{3} \sum_{k=1}^{N} \sum_{l=1}^{N}
 \left[
 b_{mk}^{(8)} ( b_{ln}^{(2)} + b_{ln}^{(1)} r_l r_n^{-1} )
 + b_{ln}^{(8)} ( b_{mk}^{(2)} + b_{mk}^{(1)} r_m r_k^{-1} )
 \right] 	\nonumber\\
 && \cdot \left[ r_l^2 r_n^{-2} \mathscr{P}_{kl}^{HH}(-{\bf r}_C)  e^{i{\bf k}\cdot{\bf r}_C}
 + r_m r_k^{-1} \mathscr{P}_{kl}^{HH}({\bf r}_A) e^{-i{\bf k}\cdot{\bf r}_A}
 + r_m^{2} r_k^{-2} r_l r_n^{-1} \mathscr{P}_{kl}^{HH}({\bf 0}) \right].
\end{eqnarray}
\end{subequations}
\end{widetext}


\begin{thebibliography}{99}

\bibitem{Lieb_Schultz_Mattis} E. Lieb, T. D. Schultz, and D. C. Mattis, Ann. Phys. {\bf 16}, 407 (1961). 

\bibitem{Cloizeaux_Pearson} J. des Cloizeaux and J. J. Pearson, Phys. Rev. {\bf 128}, 2131 (1962).

\bibitem{herbertsmithites_Helton} J. S. Helton, K. Matan, M. P. Shores, E. A. Nytko, B. M. Bartlett, Y. Yoshida, Y. Takano, A. Suslov, Y. Qiu, J.-H. Chung, D. G. Nocera, and Y. S. Lee, Phys. Rev. Lett. {\bf 98}, 107204 (2007).

\bibitem{herbertsmithites_Mendels} P. Mendels, F. Bert, M. A. de Vries, A. Olariu, A. Harrison, F. Duc, J. C. Trombe, J. S. Lord, A. Amato, and C. Baines, Phys. Rev. Lett. {\bf 98}, 077204 (2007).

\bibitem{herbertsmithites_Ofer} O. Ofer, A. Keren, E. A. Nytko, M. P. Shores, B. M. Bartlett, D. G. Nocera, C. Baines, and A. Amato, arXiv:cond-mat/0610540v2 (unpublished).

\bibitem{36-site_ED_Leung_Elser} P. W. Leung and V. Elser, Phys. Rev. B {\bf 47}, 5459 (1993).

\bibitem{36_ED_first_excited} Ch. Waldmann, H.-U. Everts, B. Bernu, C. Lhuillier, P. Sindzingre, P. Lecheminant, and L. Pierre, Eur. Phys. J. B {\bf 2}, 501 (1998).

\bibitem{ED_Sindzingre} P. Sindzingre and C. Lhuillier, arXiv:cond-mat/0907.4164v2 (unpublished).

\bibitem{Zeng_Elser} C. Zeng and V. Elser, Phys. Rev. B {\bf 42}, 8436 (1990).

\bibitem{Mila} F. Mila, Phys. Rev. Lett. {\bf 81}, 2356 (1998).

\bibitem{Sindzingre} P. Sindzingre, G. Misguich, C. Lhuillier, B. Bernu, L. Pierre, Ch. Waldtmann, and H.-U. Everts, Phys. Rev. Lett. {\bf 84}, 2953 (2000).

\bibitem{spin_liquid_Sachdev} S. Sachdev, Phys. Rev. B {\bf 45}, 12377 (1992).

\bibitem{spin_liquid_Hastings} M. B. Hastings, Phys. Rev. B {\bf 63}, 014413 (2000).

\bibitem{spin_liquid_Ran} Y. Ran, M. Hermele, P. A. Lee, and X.-G. Wen,  Phys. Rev. Lett. {\bf 98}, 117205 (2007).

\bibitem{Marston_Zeng} J. B. Marston and C. Zeng, J. Appl. Phys. {\bf 69}, 5962 (1991).

\bibitem{Nikolic_Senthil} P. Nikolic and T. Senthil, Phys. Rev. B {\bf 68}, 214415 (2003).

\bibitem{Singh_Huse_1} R. R. P. Singh and D. A. Huse, Phys. Rev. B {\bf 76}, 180407 (2007).

\bibitem{Singh_Huse_2} R. R. P. Singh and D. A. Huse, Phys. Rev. B {\bf 77}, 144415 (2008).

\bibitem{Kagome_BOT} B.-J. Yang, Y. B. Kim, J. Yu, and K. Park,  Phys. Rev. B {\bf 77}, 224424 (2008).

\bibitem{quantum_dimer_model_Poilblanc} D. Poilblanc, M.  Mambrini, and D. Schwandt, Phys. Rev. B {\bf 81}, 180402 (2010).

\bibitem{Entanglement_Renormalization_Vidal} G. Evenbly and G. Vidal, Phys. Rev. Lett. {\bf 104}, 187203 (2010).

\bibitem{JiangWengSheng} H. C. Jiang, Z. Y. Weng, and D. N. Sheng, Phys. Rev. Lett. {\bf 101}, 117203 (2008).

\bibitem{YanHuseWhite} S. Yan, D A. Huse, and S. R. White, Science {\bf 332}, 6034 (2011).

\bibitem{ED_Sorensen} A. M. L\"{a}uchli and J. Sudan and E. S. S\o rensen, Phys. Rev. B {\bf 83}, 212401 (2011).

\bibitem{ShastrySutherland} B. S. Shastry and B. Sutherland, Physica B {\bf 108}, 1069 (1981).

\bibitem{MiyaharaUeda} S. Miyahara and K. Ueda, Phys. Rev. Lett. {\bf 82}, 3701 (1999).

\bibitem{Kageyama} H. Kageyama, K. Yoshimura, R. Stern, N. V. Mushnikov, K. Onizuka, M. Kato, K. Kosuge, C. P. Slichter, T. Goto, and Y. Ueda, Phys. Rev. Lett. {\bf 82}, 3168 (1999).

\bibitem{SachdevBhatt} S. Sachdev and R. N. Bhatt, Phys. Rev. B {\bf 41}, 9323 (1990).

\bibitem{BlaizotRipka} J. P. Blaizot and G. Ripka, \emph{Quantum Theory of Finite Systems} (MIT, Cambridge, MA, 1986).

\end{thebibliography}



\end{document}